\documentclass[useAMS,usenatbib,referee]{biom}
\usepackage{amsmath}
\usepackage{hyperref}
\usepackage{todonotes}
\def\bSig\mathbf{\Sigma}

\usepackage[figuresright]{rotating}

\title[Detection of multiple perturbations in multi-omics biological networks]{Detection of multiple perturbations in multi-omics biological networks}

\author{Paula J. Griffin$^{1}$,
   W. Evan Johnson$^{1,2,4}$, and
   Eric D. Kolaczyk$^{3,4*}$\email{kolaczyk@bu.edu} \\
   $^{1}$Department of Biostatistics, Boston University School of Public Health\\
   $^{2}$Division of Computational Biomedicine, Boston University School of Medicine \\
   $^{3}$Department of Mathematics and Statistics, Boston University\\
   $^{4}$Bioinformatics Program, Boston University }

\begin{document}


\pagerange{\pageref{firstpage}--\pageref{lastpage}} 



\label{firstpage}


\begin{abstract}
Cellular mechanism-of-action is of fundamental concern in many biological studies. 
It is of particular interest for identifying the cause of disease and learning the way in which treatments act against disease.  
However,  pinpointing such mechanisms is difficult, due to the fact that small perturbations to the cell can have wide-ranging downstream effects. 
Given a snapshot of cellular activity, it can be challenging to tell where a disturbance originated. 
The presence  of an ever-greater variety of high-throughput biological data offers an opportunity to examine cellular behavior from multiple angles, but also presents the statistical challenge of how to effectively analyze data from multiple sources.  
In this setting, we propose a method for mechanism-of-action inference by extending network filtering to multi-attribute data. 
We first estimate a joint Gaussian graphical model across multiple data types using penalized regression and filter for network effects.  
We then apply a set of likelihood ratio tests to identify the most likely site of the original perturbation. 
In addition, we propose a conditional testing procedure to allow for detection of multiple perturbations.  
We demonstrate this methodology on paired gene expression and methylation data from The Cancer Genome Atlas (TCGA).
\end{abstract}

%
%


\maketitle


\section{Introduction}
\label{s:intro}
Activity within a cell is governed by a complex set of molecular interactions. 
In such an intricate system, the introduction of a perturbation to a single element in the network can have widespread effects throughout the system. 
For mechanism-of-action inference or intervention targeting, it is a critical and difficult task to distinguish the site of the original perturbation from the downstream ripple effects.
For example, testing genes one-by-one in an isolated manner, as in differential expression analyses, may be able to identify changes between two states, but the site of the largest change is not necessarily the site of an original disturbance.  
Our goal is to invert the process by which the effect propagates throughout the network, and identify the site of the initial perturbation to the system.

Previous work demonstrates the importance of considering network effects in analysis of gene expression data.  
\cite{diBernardo2005} proposed mode-of-action by network identification (MNI), which used a large microarray compendium to construct a gene interaction network, then ``filtered" expression profiles to identify the direct gene targets of each perturbation.  
Later, \cite{Cosgrove2008} provided a more statistically principled approach, SSEM-Lasso (sparse simultaneous equations model via lasso).  
This latter method consists of network estimation using lasso estimation, followed by filtering for network effects using the estimated regression parameters.
Subsequently, genes are ranked as likely perturbation sites according to the magnitude of their residuals. 
The theoretical properties of this method are explored by \cite{Yang2010}.  
Both of these methods were shown to be capable of providing improved detection of perturbation sites over methods that did not incorporate network structure, such as differential expression analysis.  
Other researchers consider this problem at the level of pathways rather than individual genes. 
\cite{Pham2011} build a pathway-level network based on differential expression and KEGG \citep{KEGG2000} pathway membership in order to identify pathways of interest.
\cite{Ma2012} pursue joint modelling in a different way, using drug sensitivity data and gene expression measurements in a Bayesian factor analysis to identify drug targets.

In addition to the difficulty of isolating the primary mover from the vast chain of trailing interactions, the recent trend of data integration introduces further modelling complexity.  
Researchers often collect measurements of multiple types on a single subject or sample, quantifying phenomena like gene expression, methylation status, and protein abundance.   
Recent efforts have established that examining a biological phenomenon from multiple `angles' using multiple types of data can provide important additional mechanistic insight \citep{Bordbar2012, Zhang2012, MacNeil2015}. 
For human studies, multiple types of measurements may be taken in order to get the most information out of a limited pool of subjects.

Though multiple measures are often collected now, the analytic techniques to cope simultaneously with multiple data types are still developing.  
In many studies, each data type is analyzed separately and then subjected to some joint postprocessing, such as a check for correlation, or annotation for proximity between sets of results (for example, \citealt{Fournier2010}; \citealt{Lee2011}; \citealt{Varambally2005}; \citealt{Tsavachidou2010}).  
Alternatively, one data type may be used as a discovery data set, while a second is reserved for validation. 
Analyses of this variety assume that there should be some mirroring of effects between data types, but typically ignore the inherent dependency between biological elements.  
For instance, the quantity of mRNA transcript is not independent of the abundance of its protein product, nor of its own methylation status. 
Various methods exist for inference of potential drug targets (for an overview, see \citealt{Lecca2013} and \citealt{Csermely2013}), but to our knowledge none have addressed the question of how to jointly model multi-type data while explicitly filtering out effects due to network-based propogation.

In this paper, we present a strategy for identifying gene-level perturbation sites in multi-type biological data.  
We construct a joint Gaussian graphical model incorporating all data types.  
Next, we estimate network structure using a graphical lasso, informed by prior data regarding gene-gene interactions.
After then filtering for network effects, we develop a ranking of likely primary perturbation sites based on a series of likelihood ratio tests.  We also
offer an extension for inference of secondary sites.  We demonstrate the efficacy of this methodology in a simulation study, and in an application to 
joint methylation and gene expression data from The Cancer Genome Atlas (TCGA; \citealt{TCGA2012}). 

\section{Joint Gaussian graphical model}
\label{s:jggm}
In defining a framework to model cellular activity, we adopt a gene-centric perspective. 
Specifically, we match attributes of $K$ different types to form a joint gene-level ``node."
We then form a graph $G=\{V, E\}$ of gene-wise interactions across these joint nodes.   
For example, a node may be constructed with a gene's $K=3$ attributes of expression, methylation status, and protein abundance.
Since we expect biologically that cross-gene interactions are relatively rare compared to interactions across measurement types, this joint-node simplification facilitates estimation, reducing the number of potential edges in $G$ from $\frac{pK(pK-1)}{2}$ to $\frac{p(p-1)}{2}$, for $p$ genes.

In more detail, for a single node $i \in \{V: 1,\ldots,p\}$, we have $K$ measurements $Y_i=[Y_i^{(1)}, \ldots, Y_i^{(K)}]^T$.  
These nodes are are combined into a ``stacked" vector $Y$ by node, writing \\ $Y=[Y_1^{(1)},Y_1^{(2)}, \ldots, Y_1^{(K)} \ldots, Y_p^{(1)},Y_p^{(2)}, \ldots, Y_p^{(K)}]^T$.
We then specify a conditional Gaussian graphical model, in which each element may be expressed as a linear combination of its neighbors, plus some perturbation $\mu$ and error $\epsilon$:
\begin{equation}\label{eq:condmodel}
y_i^{(k)}|y_{(-i)}, y_i^{(-k)} = \mu_i^{(k)} +  \sum_{l \neq k}  b_{ii}^{(k,l)} y_{i}^{(l)} + \sum_{l=1}^K \sum_{i \sim j} b_{ij}^{(k,l)} y_{j}^{(l)} + \epsilon_i^{(k)}\enskip,
\end{equation}
with $\epsilon_{i}^{(k)}\sim N(0, \sigma^2)$.  
The additional term $\mu_{i}^{(k)}$ represents an external perturbation to $Y_{i}^{(k)}$ that results in a mean-shift, and is distinct from the effects of $i$'s neighbors.
Taking all nodes jointly, we can rewrite the model of Equation~(\ref{eq:condmodel}) as  
\begin{eqnarray}\label{eq:joint}
Y &\sim& N((I-B)^{-1}\mu, (I-B)^{-1}\sigma^2)\\
Y &\sim& N(\Sigma\mu, \Sigma\sigma^2) \enskip.
\end{eqnarray}
Derivation of this formulation follows as in \cite{Cressie1993}.   
The matrix $B$ is constructed from coefficients in the conditional formulation, and so an entry $b_{ij}^{(k,l)}=0$ indicates $y_j^{(l)}$ does not directly influence $y_{i}^{(k)}$, and results in a zero in the precision matrix $\Omega=\Sigma^{-1}$.   
The vector of external perturbations $\mu$ is believed to be sparse, and our goal will be to identify likely nonzero entries in $\mu$, corresponding to perturbation sites.   

In practice, we do not know $\Sigma$, and must estimate it from our data. 
If there are no external perturbations to the network ($\mu=0$), then we have $Y \sim N(0, \Sigma)$, which allows estimation of $\Sigma$.
We define a perturbation as occurring relative to a control in case/treated data.
We assert $\mu=0$ holds in the control data, and estimate $\Sigma$ with control samples only.
We will then use $\hat{\Sigma}$ to make inferences about $\mu$ in case/treated samples.  

As the number of entries in $\Sigma$ far exceeds the available sample size, we apply a variant on the regularization of \cite{Kolar2014} in estimation of $\hat{\Sigma}$.  
For precision matrix $\Omega$, we build a block matrix according to node membership.
\begin{eqnarray}\label{eq:blockomega}
\Omega &= \left[\begin{array}{cccc}
\Omega_{11}& \Omega_{12}& \cdots & \Omega_{1p}\\
\Omega_{21}& \Omega_{22}& \cdots & \Omega_{2p}\\
\vdots & &\ddots &\vdots \\
\Omega_{p1}& \Omega_{p2}&\cdots & \Omega_{pp}\\
\end{array}\right]
\end{eqnarray}
In estimation of $\hat{\Omega}$, we apply a penalty to the Frobenius norms of these submatrices, and optimize according to 
\begin{equation}\label{eq:opt}
\hat{\Omega} = \mathrm{argmin}_{\Omega \succ 0} \left(\mbox{tr}(S\Omega) - \log|\Omega| + \lambda \sum_{a,b} w_{ab}^{-1} \|\Omega_{ab} \|_F\right)
\end{equation}

Penalizing on the level of these submatrices encourages entire $(K\times K) $ blocks in $\hat{\Omega}$ to zero.  
As previously noted, if submatrix $\Omega_{ab}=0_{K\times K}$, then nodes $a$ and $b$ are conditionally independent.  
This type of variable selection procedure is a variant of covariance selection \citep{Dempster1972}.  
Further, a zero entry in the covariance matrix $\Sigma=\Omega^{-1}$ further indicates a lack of indirect influence, meaning the nodes are in separate components of the graph $G$.
Building our network this way offers an attractive compromise between allowing interactions across data types and limiting the number of edges that must be estimated.
Optimization based on Equation~(\ref{eq:opt}) proceeds according to approximate block-gradient descent, with details in \cite{Kolar2014}.  
We recommend selection of the tuning parameter $\lambda$ based on minimum extended Bayesian information criterion with $\gamma=0.5$ \citep[EBIC;][]{Chen2008}, which we have found offers better network recovery than the Bayesian information criterion (BIC) for small sample sizes.      

In addition to the block structure, we allow an optional weight to increase the penalty on biologically unlikely edges.  
In Equation~(\ref{eq:opt}), $w_{ab}$ represents a plausibility score for between-node interactions.  
This offers biologically reasonable interactions a lower barrier to entry in the model.  
Such scores can be constructed using a database such as STRING \citep{Szklarczyk2011}, as we do in Section~4, or ENCODE \citep{Encode2004}.  
The weights may also be left at a constant value if insufficient prior information exists for the scenario at hand.   
This can facilitate estimation of larger networks with relatively few samples.

\section{Perturbation site identification}
\label{s:psi}
\subsection{Multi-attribute testing procedure}
\label{ss:matest}
Given an estimate $\hat{\Omega}$, we now proceed to our main problem of interest, i.e., inference on perturbation site in case data, through inference on $\mu$. 
\cite{Cosgrove2008} introduce the method of using an estimate of the covariance matrix to invert the propagation of network effects, which they called ``network filtering." 
We can extend this concept to multi-type data by using a joint covariance matrix, obtained by the previously outlined method. 
In order to ascertain which node has been perturbed,  we propose the use of node-wise likelihood ratio tests.  Note that,
as the material that follows in this section and the next do not depend directly on the particular choice of estimator $\hat{\Omega}$ adopted in Section~2,
we present our proposed methodology in terms of known $\Omega$ (or $\Sigma$), and then address the question of how estimation of
$\Omega$ impacts the overall procedure through a general analysis. 

For a given node $i$, we test the hypothesis that only the entries in $\mu$ corresponding to node $i$ (that is, $\mu_i = [\mu_i^{(1)}, \ldots, \mu_i^{(K)}]^T$) are nonzero ($\mu_i \neq 0$, $\mu_{(-i)}=0$), against the null hypothesis of an entirely zero mean-shift vector ($\mu=0$). 
This may be interpreted as a test of whether a particular gene has been perturbed, conditional on it being the only perturbation.   

Without loss of generality, we consider a test at the first node,  i.e., a test that $\mu_1 \neq 0$.  
We invert the network propagation and filter the data to obtain $Z= \Omega Y \sim N(\mu, \Omega)$.  
That is, through `network filtering' we produce an alternative representation of the data with mean $\mu$, rather than $\Sigma\mu$.
In this parametrization, we obtain the maximum likelihood estimator for $\mu_1$ under the alternative hypothesis as
\begin{eqnarray} \label{eqn:mle}
\hat{\mu}_1 &= \bar{z}_1 +\Sigma_{11}^{-1}\Sigma_{1 \cdot }\bar{z}_\cdot
\end{eqnarray}
where $\bar{z}_\cdot$ indicates the mean of the filtered data not being presently tested (i.e., $\bar{z}_{(-1)}$),  $\Sigma_{\cdot \cdot}$ indicates the corresponding submatrix in $\Sigma$, and so on.   
The resulting likelihood ratio test may be written
\begin{eqnarray} \label{eqn:lrtstat}
T_1 
&= n\left(\bar{z}^T\Sigma \bar{z} -  \bar{z}_\cdot^T (\Sigma_{\cdot\cdot} - \Sigma_{\cdot 1}\Sigma_{11}^{-1}\Sigma_{1\cdot}) \bar{z}_\cdot  \right) \enskip.
\end{eqnarray}

Note that the precision of the filtered data is the covariance of the data on the original scale, $\Sigma$. 
The formula for the conditional precision $Z_\cdot$ given $Z_1$ is $\mbox{Prec}(Z_\cdot | Z_1)=\Sigma_{\cdot\cdot} - \Sigma_{\cdot 1}\Sigma_{11}^{-1}\Sigma_{1\cdot}$.  
As such, the form of this test statistic is reminiscent of  Hotelling's $T^2$ statistic on the filtered data ($\bar{z}^T \mbox{Prec}(Z)\bar{z}$), less its portion deriving from the portion of $\mu$ that has been assumed-zero  ($\bar{z}_\cdot^T \mbox{Prec}(Z_\cdot | Z_1)\bar{z}_\cdot$).  
We perform this test for each node in turn, and then rank their likelihood of being the true perturbation site by test statistics $T_1, T_2, \ldots, T_p$.

Under the null hypothesis of $\mu=0$, $T_j \sim \chi^2_K(0)$ for all $j$.  
Under the the alternative hypothesis of $\mu\ne 0$, each test statistic $T_j$ has a noncentral chisquare distribution.  
For example, for $j=1$, this takes the general form
\begin{eqnarray}\label{eq:chisqdist}
T_1 &\sim&  \chi^2_K \left( \mu^T \left(\begin{array}{cc}
\Sigma_{11} & \Sigma_{1\cdot} \\
\Sigma_{\cdot 1} & \Sigma_{\cdot1}\Sigma_{11}^{-1}\Sigma_{1\cdot}\\
\end{array}  \right)\mu \right) \enskip .
\end{eqnarray}
Suppose that the true perturbation is located at the first gene, i.e., that $\mu_1\ne 0$ and $\mu_\cdot = 0$.  
Comparing $T_1$ with a test at another node $j\ne 1$, we obtain
\begin{eqnarray}
T_1 &\sim& \chi^2_K (\mu_1^T \Sigma_{11}\mu_1)\\
T_j &\sim& \chi^2_K (\mu_1^T \Sigma_{1 j} \Sigma_{jj}^{-1} \Sigma_{j1} \mu_1)  \enskip .
\end{eqnarray}
Since $\Sigma_{11}-\Sigma_{1 j} \Sigma_{jj}^{-1} \Sigma_{j1}$ is positive-definite, $(\mu_1^T \Sigma_{11}\mu_1) > (\mu_1^T \Sigma_{1 j} \Sigma_{jj}^{-1} \Sigma_{j1} \mu_1)$, and $T_1$ stochastically dominates $T_j$ for any node $j$ not containing a true perturbation.

While these derivations are shown here as a node-wise test, this test can be applied to any predefined sets of nodes of arbitrary size and overlap. 
In principle, testing could be based on individual elements of $\mu$, or on entire pathways.  
The test statistics $T$ may not be directly compared if groups of varying sizes are tested, but  $p$-values may be calculated on the basis of the chisquare distribution, with degrees of freedom equal to the total number of nodes in the group being tested.  

\subsection{Sequential multi-target testing}
\label{ss:mttest}

We have so far considered the occurrence of a single perturbation, but this is not always realistic.  
A treatment may have off-target effects, resulting in multiple interaction sites \citep{Afzal2014}, or a disease may be caused by perturbations to more than one gene.  
In such a case, interpretation of the previously described results becomes less straightforward. 
Since each of our previously described tests assumes that all other nodes have zero mean, we automatically perceive nodes {\it near} the truly perturbed node to be likely sites, so a near-target effect may be confused with a distinct, off-target effect.  
Once we have identified a primary perturbation site, we may wish to consider the most likely site for a secondary perturbation, in a manner that accounts for
the location of the first.  

Nested likelihood ratio tests provide a natural framework for a sequential ranking. 
At step $s+1$, we denote the sites already identified in steps $1, \ldots, s$ as a set $S$.
Having already determined that that the subvector $\mu_{S}$ of $\mu$ contains nonzero entries, we can conduct a likelihood ratio test on the remaining nodes to search for additional perturbations.   
Thus, at step $s+1$, for node $i$, we test the hypothesis that an additional perturbation is located at node $i$ ($\mu_i \neq 0, \mu_{S} \neq 0, \mu_{-(S,i)} = 0$) against the null that no perturbations outside of $S$ exist ($\mu_{S} \neq 0, \mu_{-(S)} = 0$).   
We perform this calculation for all nodes $i$ not determined to be perturbation sites in steps $1, \ldots, s$.  

The resulting test statistic  $T_i ^{[s+1]}$ may be written as a difference of unadjusted likelihood ratio test statistics:
\begin{eqnarray}
 T_i^{[s+1]} = T_{(i,S)} - T_{S} \enskip,
\end{eqnarray}
where  $T_{S}$ corresponds to testing $\mu_{S} \neq 0$, $\mu_{(-S)}=0$ against $\mu=0$, and $T_{(i,S)}$ corresponds to testing $\mu_i \neq 0$, $\mu_S \neq 0$, $\mu_{(-i,-S)}=0$ against $\mu=0$,
Inference can proceed on the conditional sequence, or  $p$-values can be calculated and adjusted to maintain an appropriate false discovery rate across $s$ using the method of \cite{Benjamini2001}.  

The magnitude and direction of the difference between this value and the original test statistic depends upon the correlation between the node currently being tested and the nodes already ``found'' by the sequential procedure.  Theorems~\ref{th:expnest} and \ref{th:nested} establish some properties relevant to the relative ranking of the adjusted test statistics.

\begin{theorem}{\label{th:expnest}}
Given a set of nodes already found to have nonzero mean in steps $1, \ldots s$, consider testing for a perturbation at an additional node $i$ in step $s+1$.  Denote the indices in $Z=\Omega Y$ corresponding to the nodes found in steps $1, \ldots, s$ as $S$. 

We can write the expected difference between the original test statistic and the test statistic adjusted for perturbations in $S$ as 
\begin{equation*}
E(T_i - T^{[s+1]}_i) = \mu_i^T\left(\Sigma_{i,S}\Sigma_{S,i} \right) \mu_i  + 
						 2 \mu_i^T \left(\Sigma_{i,S} \right) \mu_S + 
						 \mu_S^T \left(\Sigma_{S,i} \Sigma_{i,S} \right) \mu_S
						\enskip .
\end{equation*}
In the special case that $\mu_i = 0$, 
\begin{align*}
E(T_i - T^{[s+1]}_i | \mu_i = 0 ) \geq 0 \enskip.
\end{align*}

\end{theorem}

\noindent As such, if no perturbation is truly present at node $i$, we expect its adjusted test statistic to be no larger than the unadjusted statistic.  

\begin{theorem}{\label{th:nested}}
Under the same conditions outlined in the general case of Theorem~\ref{th:expnest}, if $\Sigma_{i,S} = 0$, then
\begin{equation*}
T^{[s+1]}_i= T_i \enskip .
\end{equation*}
\end{theorem}

The proofs of Theorems~\ref{th:expnest} and \ref{th:nested} are given in Supplementary Materials, Section 2.
Taken together, these facts give us insight into the way that secondary targets are identified. 
Suppose we test for secondary perturbations at nodes $i$ and $j$ after finding an initial set of nodes $S$. 
When $\Sigma_{i,S}=0$, $i$ and $S$ are not connected in our graph, and the sequential test statistic for $i$ is the same as the unadjusted statistic.
Simultaneously, a correlation between measurements on $j$ and $S$ removes the near-target effects due to proximity to $S$, resulting in an expected decrease in $T_j^{[s+1]}$ compared to $T_j$ by  $\mu_S^T \Sigma_{S,j}\Sigma_{j,S} \mu_S$ . 
Since at any step $s$ we are concerned with relative ranking of test statistics, the decreased $T_j^{[s+1]}$ relative to $T_i^{[s+1]}$ makes $i$ a better candidate for an additional perturbation than it was previously.
Accordingly, this  procedure has the largest potential benefit when the two perturbations are completely separated in the graph.

\begin{figure}[t]
\centerline{
\includegraphics[width=.32\linewidth]{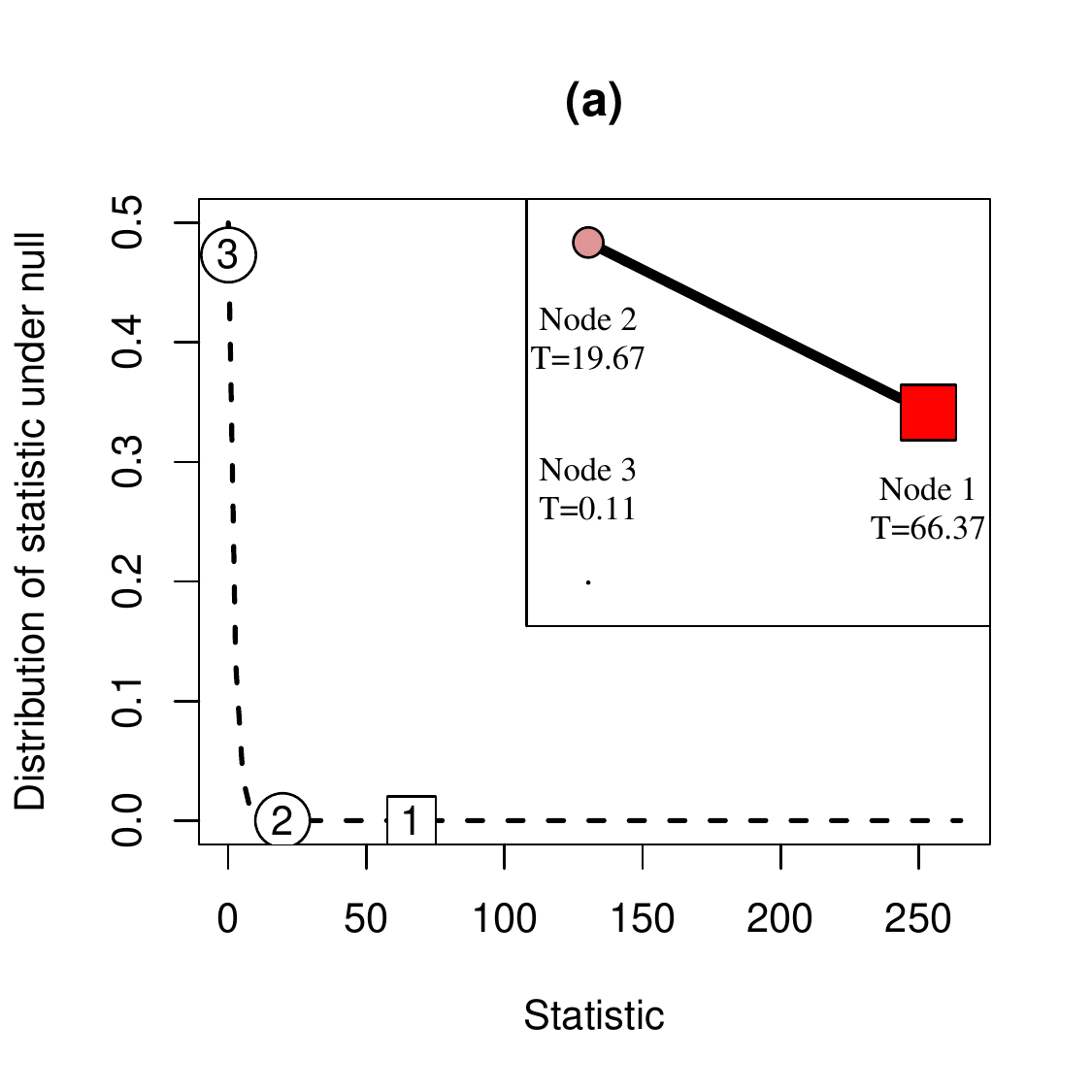}
\includegraphics[width=.32\linewidth]{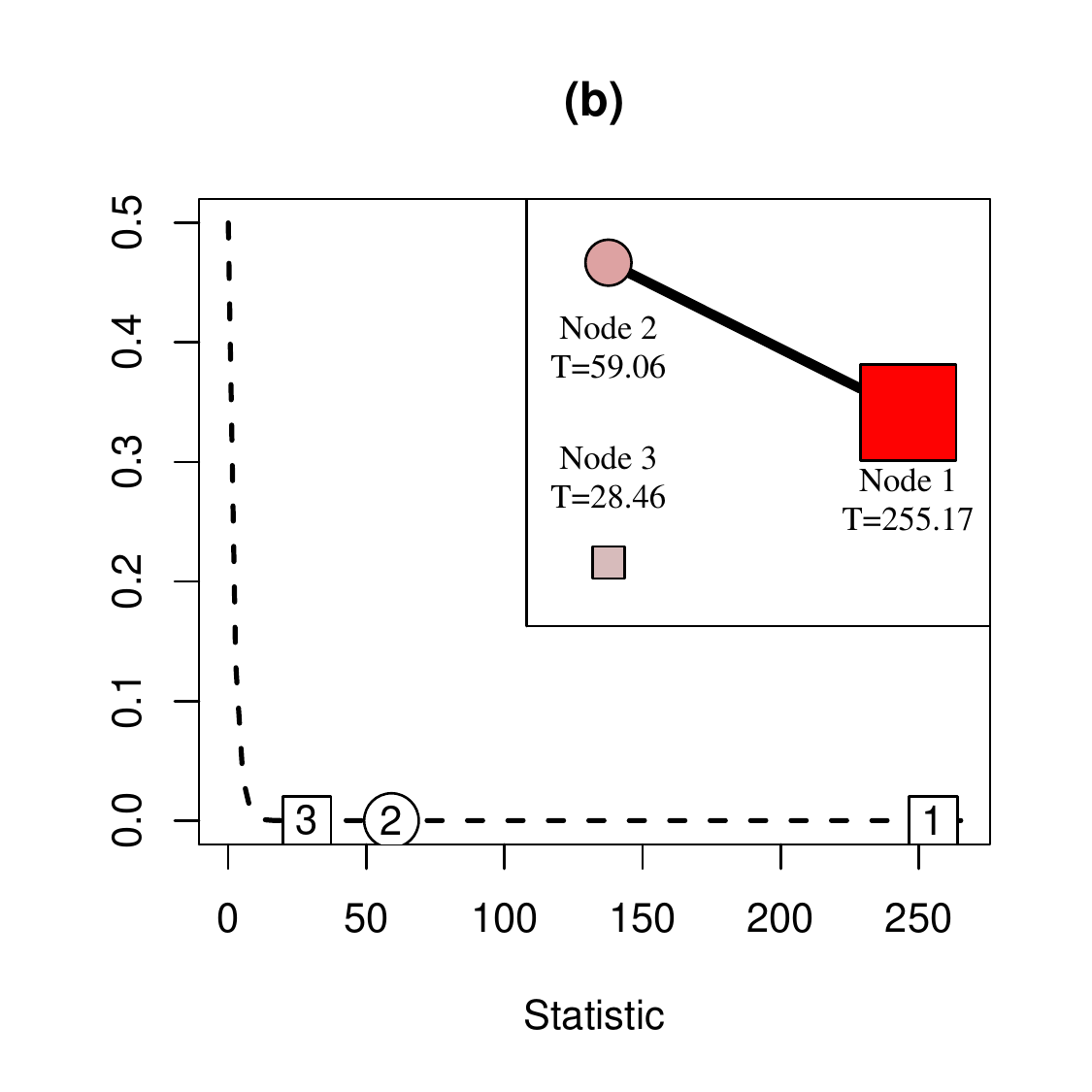}
\includegraphics[width=.32\linewidth]{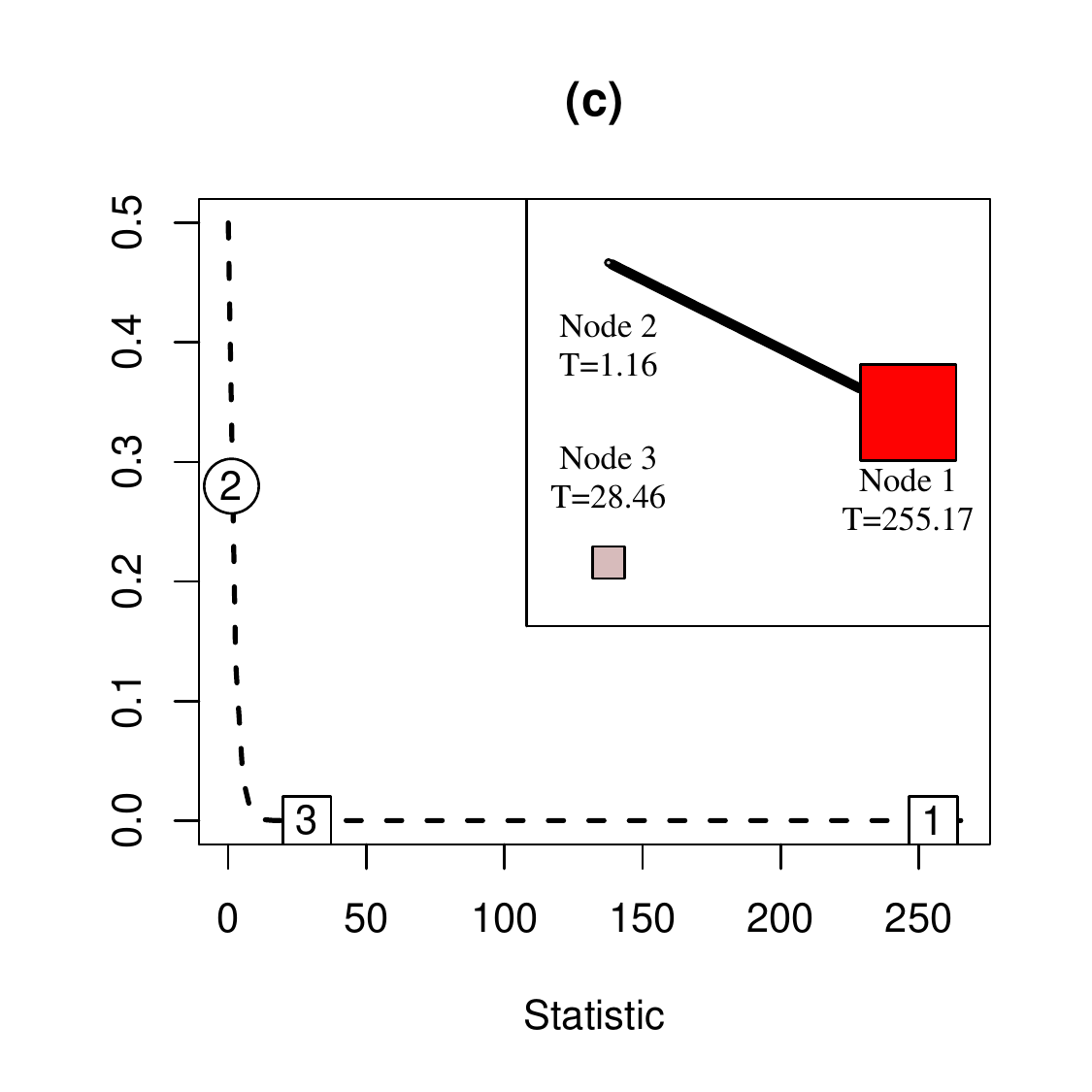}}
\caption{A toy example illustrating the properties of the multi-attribute NF in a 3-node network.   
Perturbed nodes are shown as squares, and node area is representative of test statistic size.  
Nodes 1 and 2 are neighbors. 
\textit{(a)} Node 1 is perturbed.  
As a neighbor to the perturbed node, 2 is identified as the second most likely site for a perturbation if only one exists.
\textit{(b)} Nodes 1 and 3 are perturbed, and the multi-attribute network filtering (NF) is applied.  
Node 2 is identified as the second most likely perturbation site because of the shared edge with node 1. 
\textit{(c)} As in (b), nodes 1 and 3 are perturbed, but the sequential NF procedure is applied.  
After conditioning on node 1, node 3 is identified as the most likely site for a second perturbation.}\label{fig:toy}
\end{figure}

For an illustration, see Figure~\ref{fig:toy}.  
This simple network of $n=100$ samples has only $p=3$ nodes, each with $K=2$ attributes, and a single edge between nodes 1 and 2. 
In $\Omega$, we set the within-node partial correlation $\rho_{in}$, to $0.8$ and the between-node partial correlation $\rho_{out}$ to $0.2$.
In Figure~\ref{fig:toy}(a), only a single perturbation is present, at node 1, with signal-to-noise ratio (the value of the perturbation size of $\mu$ relative to a diagonal element of $\Omega$) $\mbox{SNR}=1$. 
Node 1 is ranked as the most likely perturbation site, followed by node 2.  
This is desirable behavior in \ref{fig:toy}(a) -- if we know that only one perturbation exists, then node 2 is the next-best choice.  
In \ref{fig:toy}(b,c), we add a second perturbation at node 3 with a weaker signal ($\mbox{SNR}=0.25$).   
According to the initial multi-attribute network filtering (NF) ranking shown in \ref{fig:toy}(b), node 2 is the runner-up due to its proximity to node 1.  
However, if we condition on the presence of a perturbation at node 1 as in \ref{fig:toy}(c), then node 3 is considered a more likely site for a {\it second} perturbation than node 2.  

Performance of the sequential procedure is discussed in Section~\ref{ss:multisim}.

\subsection{Accuracy}
\label{ss:accuracy}
We have described our proposed procedure for detecting multiple perturbation sites in multi-omics data 
as if the precision $\Omega$ (or covariance $\Sigma$) were known.  In practice, of course, to expect
exact knowledge of $\Omega$ is unrealistic.  Firstly, error in estimation may occur.  
In addition, we take the network estimated in the control data to be representative of the network in the case/treated data,
but if the network itself is dysregulated, this may not be an appropriate assumption.  While a detailed practical
examination of these various sources of errors and their impact on our procedure is beyond the scope of this paper, 
we provide here a general characterization result.

Without loss of generality, let $\sigma^2=1$ and consider the case of $T_j$ for $j=1$.
Let $\tilde\Omega = \Omega + \Delta$ be an erroneous version of the true $\Omega$, 
and denote by $\tilde T_1$ the corresponding version of $T_1$ resulting from 
using $\tilde\Omega$ in place of $\Omega$.  Our interest will be on the distribution of the discrepancy
$T_1 - \tilde T_1$.  Towards that end, we define the $K\times K$ matrix
$$D = \Omega_{11} - \Omega_{1\cdot}\Omega_{\cdot\cdot}^{-1}\Omega_{\cdot 1} -
         \left(\tilde\Omega_{11} - \tilde\Omega_{1\cdot}\tilde\Omega_{\cdot\cdot}^{-1}\tilde\Omega_{\cdot 1} \right) \enskip .$$
Assume $\Sigma_{11}$ is positive definite.  For the product $D\Sigma_{11}$, express its spectral decomposition as
$$D\Sigma_{11} = \sum_{k=1}^s a_k E_k\enskip ,$$
such that $\hbox{rank}(E_k) = r_k$ (corresponding to the multiplicity of the eigenvalue $a_k$) and $\sum_{k=1}^s r_k = K$.

We then have the following result.
\begin{theorem}{\label{th:errbound}}
Under the conditions above, the discrepancy $T_1-\tilde T_1$ is equal in distribution to a linear combination
of mutually independent, noncentral chisquare random variables,  
\begin{equation}
\sum_{k=1}^s a_k \chi^2_{r_k}\left(\delta_k\right) \enskip ,
\label{eq:main.decomp}
\end{equation}
where 
$$\delta_k = (n/2) \mu^T \Sigma_{\cdot 1} E_k \Sigma_{11}^{-1} \Sigma_{1\cdot}\mu \enskip .$$
Accordingly,
\begin{equation}
E\left[ T_1 - \tilde T_1\right] = tr\left(D\Sigma_{11}\right) + \frac{n}{2}\mu^T \Sigma_{\cdot 1} D\Sigma_{1\cdot} \mu
\label{eq:mean}
\end{equation}
and
\begin{equation}
\hbox{Var}\left(T_1 - \tilde T_1\right) = 2 tr\left( (D\Sigma_{11})^2\right) 
             + 2n \mu^T \Sigma_{\cdot 1} D \Sigma_{11} D \Sigma_{1\cdot} \mu \enskip .
\label{eq:var}
\end{equation}
\end{theorem}
The proof of this theorem is given in Supplementary Materials, Section 3.
The distributional result follows from application of Theorem 1 of \cite{baldessari1967distribution}, while the moment results follow from definition of first second and moments of noncentral chisquare random variables.
In the case that $\Sigma_{11}$ is not positive definite, more general results in
\cite{tan1977distribution} may be used, at the cost of additional notation and conditions. 
  
Note that $D$ in our results above, as a function of 
$\Delta = \tilde\Omega - \Omega$, plays the key role of 
capturing the impact of the discrepancy between $\Omega$ and $\tilde\Omega$.
A more relaxed -- but arguably more informative -- statement of our moment
results is the following, wherein the role of $\Delta$ is made explicit.
\begin{corollary}{\label{cor:meanvar}}
Let $||\cdot ||_2$ denote the spectral norm.  Then
$$E\left[T_1 - \tilde T_1\right] = O\left(||\Delta||_2\right) \quad\hbox{and}\quad
\hbox{Var}\left(T_1 - \tilde T_1\right) = O\left(||\Delta||^2_2\right) \enskip .$$
\end{corollary}
Hence, we see that for a given discrepancy $\Delta$ between the true $\Omega$ and the value $\tilde\Omega$, 
the expected level of discrepancy between the corresponding statistics $T_1$ and $\tilde T_1$, as well as
the standard deviation, are both of magnitude on the order of the spectral norm of $\Delta$. 
Proof of the corollary may also be found in Supplementary Materials.

\section{Simulation}
\label{s:sims}
\subsection{Single-target simulations}
\label{ss:singlesim}
We want to consider two aspects of potential performance gains: (1) conducting a network-aware analysis method, and (2) using multiple data sources.  
To our knowledge, no other method has yet been proposed for joint modeling and detection of perturbations in this multi-attribute setting.  
As such, we conduct comparisons in simulation against established methods for single-type data, and a na{\"i}ve extension of these methods to accommodate multi-type data.
To assess gains from network analysis, we compare our method with simple differential expression ($t$-tests for single-attribute data, and Hotelling's $T^2$ for multi-attribute).  
To examine the benefit from considering multiple data sources, we consider the improvement obtained from using $K=2$ sources, versus a single data type.  
We also perform SSEM-Lasso \citep{Cosgrove2008} for the single-attribute case.

We simulate data across a range of network conditions, varying the strength of associations between data types and nodes.  
We construct a network of $p=20$ nodes according to a stochastic block model \citep{Holland1983}, with $n=50$ cases and controls.  
The network is divided into two groups of nodes, where cross-block connections are more likely to occur within a block (probability $\theta_{within}=0.4$) than between blocks (probability $\theta_{across}=0.2$).
Network links are assigned $-\rho_{out}$ in the precision matrix. 

For each node with $K=2$ attributes, we first assign all within-node correlations the value $-\rho_{in}$ in the precision matrix, creating a block-structure along the diagonal.  
A small value is added to the diagonal of $\Omega$ until the minimum eigenvalue is at least $0.5$ to ensure invertibility, then the precision matrix is scaled to have diagonal $1$. 
For each network constructed, for node $i$ to be perturbed means that a mean-shift $\mu_i$ is applied to its elements.
We simulate null data from $N(0, \Sigma)$ and perturbed data with one nonzero node in $\mu$ from $N(\Sigma\mu, \Sigma)$, and perform the aforementioned estimation and testing procedure.

\begin{figure}[ht]
\centerline{\includegraphics[width=1\textwidth]{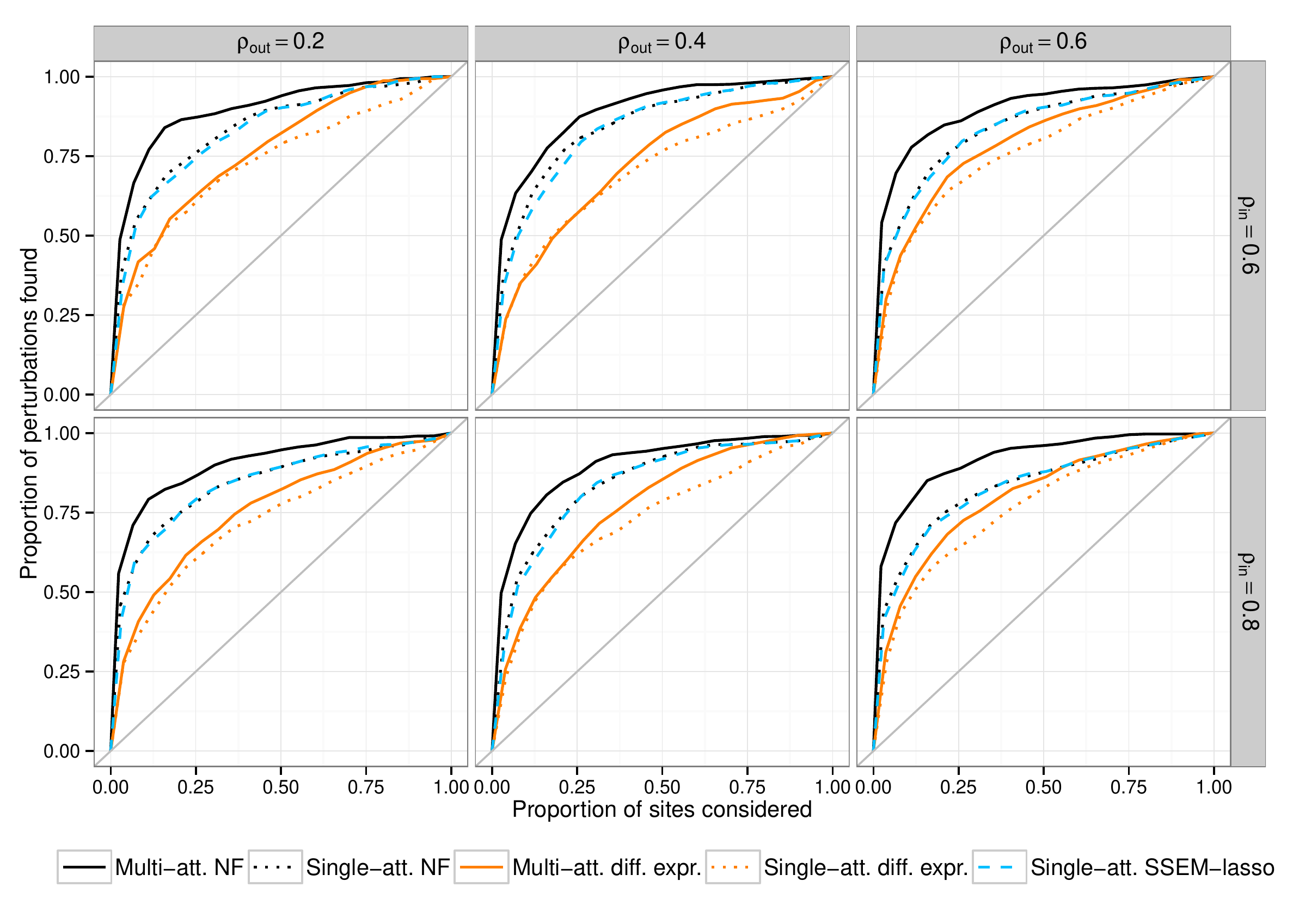}}
\caption{Single-site recovery from a stochastic block model simulation with $p=20$ nodes, $n=50$ cases and controls, and $\mbox{SNR}=0.20$. Along the $x$-axis, we consider the proportion of all sites in a top $k$ list, and along the $y$-axis, the probability that the truly perturbed site is contained within that top $k$ list.  In each plot, the jump at the leftmost edge of the graph corresponds to the probability of identifying the true perturbation as the highest-ranked site (values in Table~\ref{tab:singlesim}).  }\label{fig:singlesim}.
\end{figure}

From the likelihood ratio tests, we obtain a ranked list of nodes, with our truly perturbed node sitting at rank $r$. 
For each of $100$ simulated networks, we perturb each of the $p=20$ nodes in turn and observe their rank according to the multi-attribute network filtering (NF) procedure.  
We average over the proportion of sites occurring in our ranked list and construct receiver-operator characteristic (ROC) curves.  
These curves can be directly related to an empirical CDF, with positions along the $x$-axis indicating the proportion of total sites in a top $k$ list.  
The $y$-axis, then, indicates the probability that the true perturbation site was included in that list of $k$ sites.  Results for single-perturbation simulations are shown in Figure~\ref{fig:singlesim}.  
In addition, the probability that the top-ranked site correctly identifies the perturbation is shown in Table~\ref{tab:singlesim}.

\begin{table}[ht]
\caption{Probability that the top-ranked site is the true perturbation site and (AUC) for simulations shown in Figure~\ref{fig:singlesim}. $\rho_{in}$ indicates the strength of within-node partial correlation, and $\rho_{out}$ of cross-node partial correlations.  }
\label{tab:singlesim}
\begin{tabular}{ccccccc}
  \Hline
 &  & \multicolumn{2}{c}{NF methods}  & \multicolumn{2}{c}{Differential expression} & SSEM-lasso \\ 
$\rho_{in}$ & $\rho_{out}$ & Multi-att. & Single-att. & Multi-att. & Single-att.  & Single-att.\\ 
  \hline
  0.8 	& 0.2 & 0.56 (0.90) & 0.46 (0.84) & 0.28 (0.76) & 0.28 (0.72) & 0.41 (0.84) \\ 
   		& 0.4 & 0.50 (0.90) & 0.35 (0.84) & 0.26 (0.77) & 0.23 (0.73) & 0.32 (0.84) \\ 
   		& 0.6 & 0.58 (0.92) & 0.44 (0.83) & 0.31 (0.80) & 0.28 (0.76) & 0.41 (0.83) \\ \hline
  0.6 	& 0.2 & 0.49 (0.89) & 0.39 (0.84) & 0.28 (0.76) & 0.28 (0.73) & 0.34 (0.83) \\ 
   		& 0.4 & 0.49 (0.89) & 0.37 (0.84) & 0.24 (0.73) & 0.23 (0.70) & 0.34 (0.84) \\ 
  		& 0.6 & 0.54 (0.90) & 0.41 (0.84) & 0.30 (0.79) & 0.27 (0.76) & 0.40 (0.83) \\ 
   \hline
\end{tabular}
\vspace{1em}
\end{table}

Across a range of partial correlations, multi-attribute network filtering (NF) has most successful recovery of the perturbed site with respect to AUC and the probability of selecting the true perturbation as the top-ranked site (an ``ideal detection"). Multi-attribute NF is followed by its single-attribute counterpart and SSEM-lasso.  
Hotelling's $T^2$ follows, narrowly but consistently outperforming standard differential expression on a single attribute. 
Under all correlation settings considered here, the multi-attribute modeling strategy identifies the site correctly more than half of the time.  
On average, such ideal detections are made 54.0\% of the time for multi-attribute NF, 42.8\% for its single-attribute counterpart.  
By contrast, differential expression ranks the truly perturbed site first only 27.0\% of the time using either method.  
SSEM-lasso with a single attribute identifies the true perturbation first 39.3\% of the time, despite a comparable AUC to the single-attribute NF method, as shown in Table~\ref{tab:singlesim}.

\subsection{Multi-target simulations}
\label{ss:multisim}
We also wish to evaluate the performance of the sequential procedure when multiple perturbations are present.  
As previously noted, any advantage over simply taking the initial rankings will depend upon the network structure and the distance between perturbations.  
If two perturbations occur adjacent to one another, the near-target and off-target effects will be aligned, and the ranking will not be substantively changed.  
However, if the perturbations are far apart in the graph, this procedure may substantially improve the chances of detecting both effects.  

\begin{figure}[ht]
\centering
\includegraphics[width=1\textwidth]{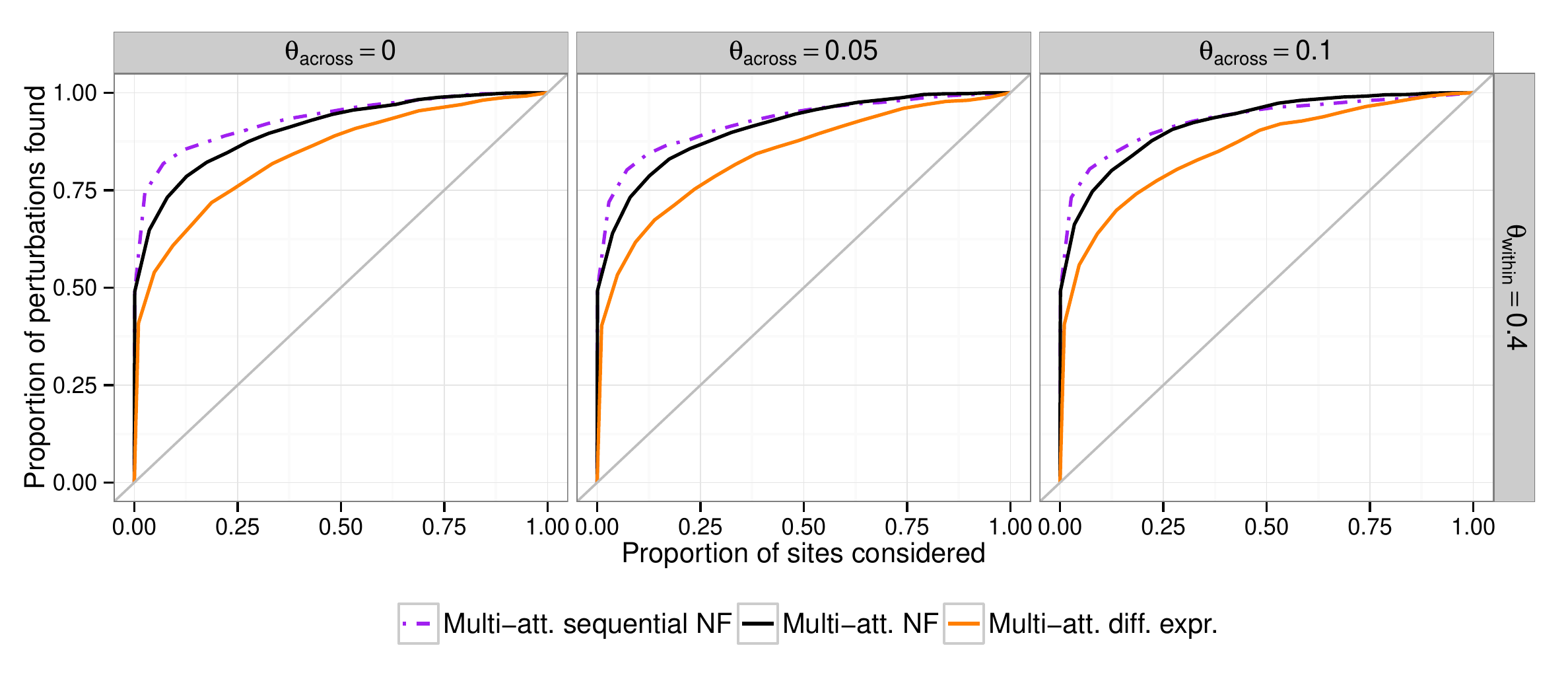}
        \caption{Simulations showing improvement of the sequentially restricted NF procedure versus the standard multi-atrribute NF and Hotelling's $T^2$ ranking when two perturbations are present, located in different blocks in a  stochastic block model.  
        The expected distance between these two perturbations are on the graph is determined by $\theta_{across}=(0,0.05, 0.1)$, corresponding to complete, moderate, and slight separation between the two blocks, relative to the within-block edge probability of $0.4$. 
        Benefits from the sequential procedure are largest when the two perturbations are not connected in the graph (left).\label{fig:multi}}
\end{figure}

We extend our previous simulations study to include a second perturbation.  
In the context of a stochastic block model, we simulate two perturbations: a nonzero node in the first block with $\mbox{SNR}=0.20$ as before, and a second, weaker perturbation in the second block with $\mbox{SNR}=0.10$.  
We then vary the probability of a cross-block edge $(\theta_{across})$ relative to the probability of an edge within each block $(\theta_{within})$ to demonstrate the role of distance on the graph in the efficacy of the sequential procedure.  
We consider $\theta_{across}/\theta_{within} = 0.25$ (slight separation), $0.125$ (moderate separation), and $0$ (complete separation). 
Table~\ref{tab:first2} shows the probability of ranking both true perturbations in the top two sites, and Figure~\ref{fig:multi} shows the ROC curves for identifying both perturbations.  
The sequential procedure outperforms the initial ranking on both counts for these scenarios, with gains increasing according to separation between the perturbations for probability of ideal identification. 

\begin{table}[ht]
\centering
\caption{Probability of identifying the both truly perturbed sites in the first two ranked positions and (AUC), considering only multi-attribute methods. Corresponding plots are shown in Figure~\ref{fig:multi}.\label{tab:first2}}
\begin{tabular}{c|ccc}
  \Hline
$\theta_{across}/\theta_{within}$ & Sequential multi-att. NF &  Multi-att. NF & Multi-att. diff. expr. \\ 
  \hline
  0.250 & 0.74 (0.93) & 0.67 (0.92) & 0.57 (0.85)\\ 
  0.125 & 0.73 (0.93) & 0.65 (0.91) & 0.54 (0.84)\\ 
  0.000 & 0.76 (0.93) & 0.66 (0.91) & 0.55 (0.84) \\ 
   \hline
\end{tabular}
\vspace{1em}
\end{table}

In certain circumstances, the sequential procedure may produce suboptimal results.  
For example, suppose that the first identification is a false positive due to proximity to a true perturbation.  
The truly perturbed site will have a lower ranking after conditioning for the false positive site, as this procedure would adjust away some of that node's own signal.  
This is particularly likely to occur when signal-to-noise ratio is low, or when multiple perturbations have common neighbors.  
As such, we recommend the use of this procedure when an unambiguous initial identification has been made, and suspected secondary perturbations are not in close proximity to the initial site.

\subsection{Comparison to post-analysis aggregation}
\label{sec:postagg}
\begin{figure}[ht]
\centering 
\includegraphics[width=1\textwidth]{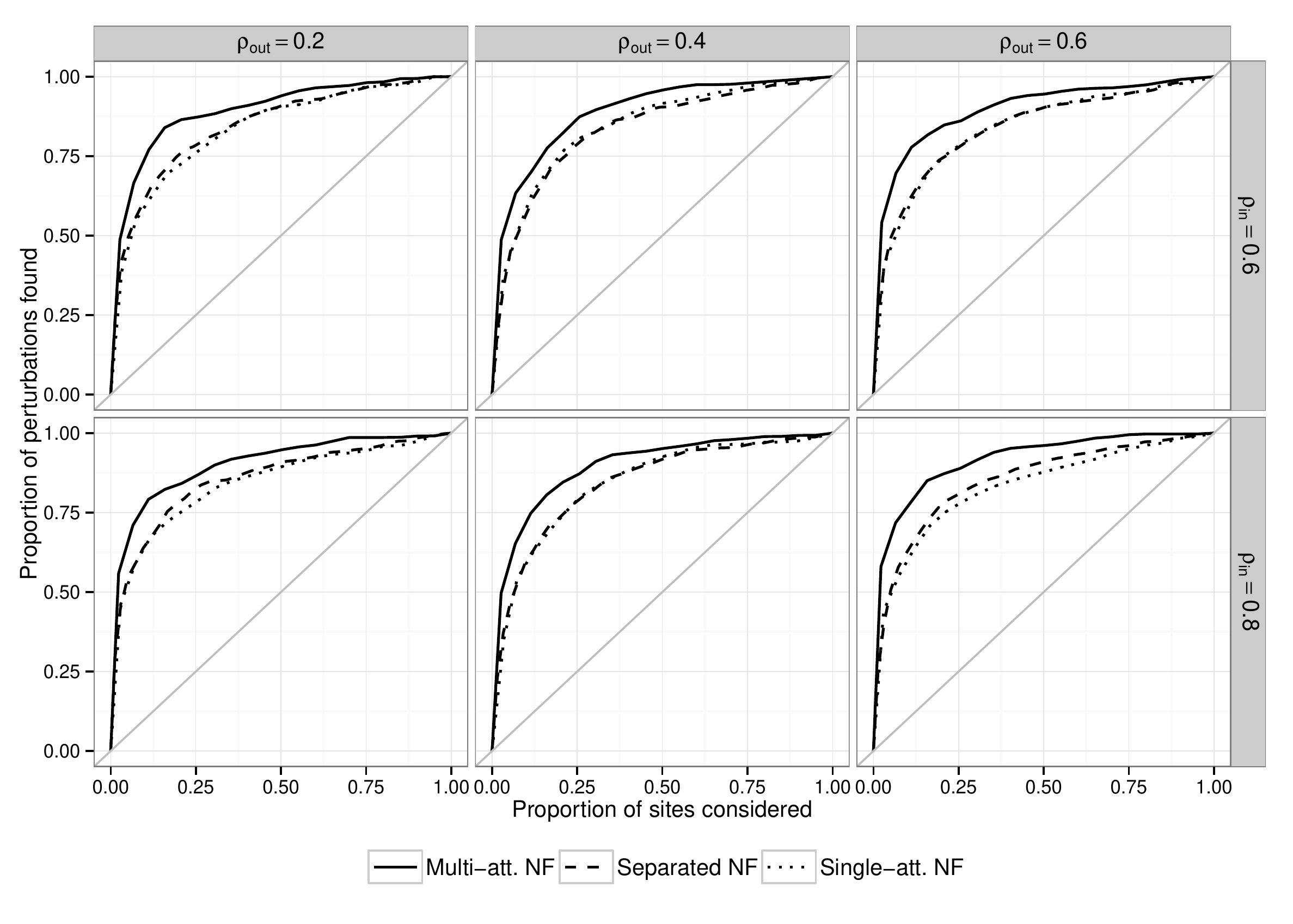}
\caption{Comparison of network filtering methods in a single-perturbation setting.   
ROC curves show perturbation site recovery from a stochastic block model simulation scheme with $p=20$ nodes, $n=50$ cases and controls, and $\mbox{SNR}=0.20$. 
Separated NF indicates that the network estimation and filtering procedures were performed in isolation on each data type and then combined for ranking. \label{fig:simsep} }
\end{figure}

While the multi-attribute NF method provides improved perturbation site detection over single-attribute methods and multivariate differential expression, we wish to consider how much is gained by considering cross-attribute relationships, as opposed to some comparatively simpler `aggregation' of single-attribute results.
This benchmark is of particular interest given the popularity of network recovery methods by \cite{Guo2011} and \cite{Danaher2013} for simultaneous inference of multiple, related networks across data types, but without cross-type interactions.  
Following the same simulation strategy as described in Section~\ref{ss:singlesim}, we consider the performance of a ``separated" ranking procedure, in which we estimate and filter for separate networks for each data type, then combine results into a block-precision matrix to rank individual biological attributes, setting cross-type entries to zero.  
This amounts to asserting independence between each data type.
Results are shown in Figure~\ref{fig:simsep}, and Table~\ref{tab:sim2}. 
Note that for the separated procedure, we look for the probability that both attributes of the perturbed node are ranked highly.  

\begin{table}[ht]
\centering 
\caption{Probability that the top-ranked sites are the truly perturbed gene and (AUC) for simulations shown in Figure~\ref{fig:simsep}.  These simulations feature a single perturbation.\label{tab:sim2} }\vspace{.5em}
\begin{tabular}{cc|ccc}
\Hline
&  & \multicolumn{3}{c}{NF methods} \\ 
$\rho_{in}$ & $\rho_{out}$ & Multi-att. & Separated & Single-att. \\ 
  \hline
0.8 & 0.2 & 0.55 (0.90) & 0.44 (0.86) & 0.41 (0.84) \\ 
    & 0.4 & 0.56 (0.90) & 0.40 (0.84) & 0.41 (0.84) \\ 
    & 0.6 & 0.46 (0.92) & 0.37 (0.85) & 0.42 (0.83) \\ \hline
0.6 & 0.2 & 0.57 (0.89) & 0.44 (0.85) & 0.48 (0.84) \\ 
    & 0.4 & 0.55 (0.89) & 0.41 (0.83) & 0.44 (0.84) \\ 
    & 0.6 & 0.55 (0.90) & 0.38 (0.84) & 0.41 (0.84) \\ 
  \hline
\end{tabular}
\vspace{1em}
\end{table}

The multi-attribute NF performs best in terms of AUC and the probability of ideal identification.  
Separated and single-attribute methods perform comparably to each other by both of these metrics.  
This also holds if we rank according to the first appearance of a gene's measurements, rather than requiring top ranks for both.  
Given that a slightly higher burden is imposed on the separated method than the single-attribute (two attributes must be ranked highly rather than one), this is a slight advantage to the separated method over analysis of a single attribute. 
Nevertheless, our results indicate that most benefits attained from this type of data integration emerge from consideration of interaction {\it between} attributes when such interactions are present in the underlying data. 
The design of our model specifically exploits the existence of cross-type interactions, and is able to better discover perturbation sites as a result.

\section{Analysis of TCGA breast cancer data}
We apply this methodology in an analysis of breast cancer data from The Cancer Genome Atlas (TCGA).
We have gene expression and methylation data obtained from tumor samples of $60$ patients with metastatic cancer and and $569$ with nonmetastatic cancer.
Both the expression and methylation data were downloaded as Level 3 normalized data, and then processed to achieve approximately Gaussian distributions. 
RNA-seq data was preprocessed by TCGA using RSEM \citep[RNASeq
by Expectation Maximization; ][]{Li2011} and MapSplice \citep{Wang2010}.  
Transcripts per million (TPM) were then transformed via quantile normalization on  $\log_2(\mbox{TPM}+1)$.
The 450k methylation array data was preprocessed by TCGA using the ratio of the intensity of methylated probes to the total probe intensity to produce $\beta$ values \citep{Du2010}.
We then transformed these values according to $\log_2 \left(\frac{\beta}{1-\beta} \right)$.
For our analysis, we extracted measurements from 274 genes up- or down-regulated in an analysis of TGF-$\beta$-mediated cancer progression of hepatocytes performed by \cite{Gotzmann2006}.  
If more than one measurement was present per gene attribute (multiple methylation sites or transcript segments), a $90\%$ trimmed mean was taken.  
Subjects were considered to have metastatic cancer if classified as such at baseline or at any subsequent follow-up.
Details of the data processing may be found in the Supplementary Materials.

We first estimate the block-precision matrix of the network using the $n=569$ tumor samples from nonmetastatic cancers.  
Using our estimated precision matrix $\hat{\Omega}$, we filter for network effects in the data from $n=60$ metastatic cases, and perform gene-wise likelihood ratio tests in order to ascertain which gene is the most perturbation candidate.  

\begin{figure}[ht]
\centering
\includegraphics[width=.7\linewidth]{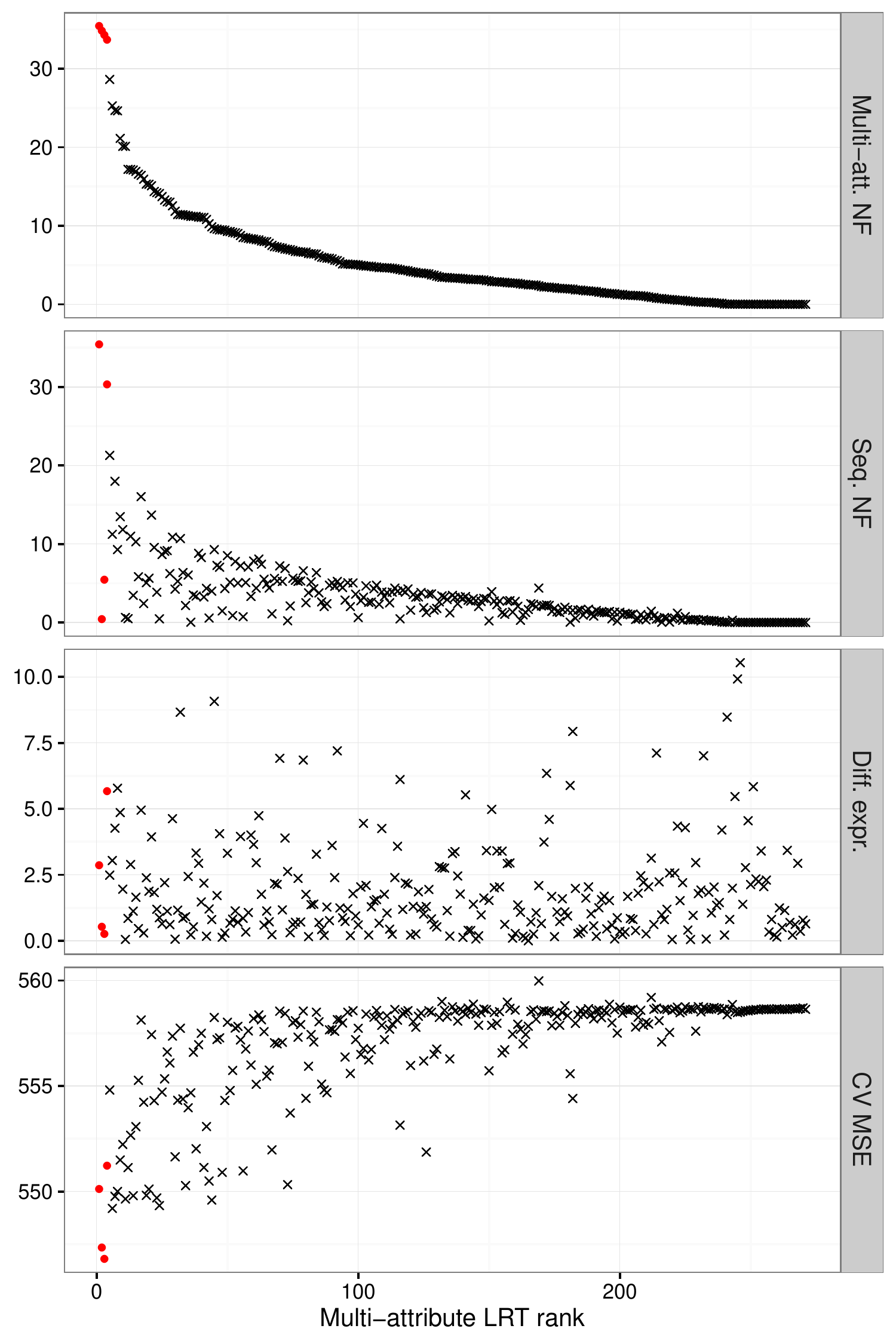}
\caption{Results from an analysis of data from TCGA.  Rank according to the non-sequential NF ranking is shown along the $x$-axis for all plots. 
Panels show NF statistic, cross-validation MSE, and differential expression statistic. 
The top 4 results  shown in Table~\ref{tab:tcga10} are highlighted in red. \label{fig:tcgastats}}
\end{figure}

The top-ranked sites are shown in Table~\ref{tab:tcga10}.
The node-wise Benjamini-Yekutieli adjusted p-values are significant  at $p_{adj}<0.05$ for 18.1\% of the genes in our list, a strong indication that at least one perturbation is present  \citep{Benjamini2001}.   
A drop-off in the test statistic is visible after the 4th position (for \textit{IRGM}, $p_{adj}=3.22e-06$, while the next gene \textit{MMP13} has $p_{adj}=3.27e-05$). 
This drop-off is visible in the top panel of Figure~\ref{fig:tcgastats}.  
As such, we consider the top 4 genes in Table~\ref{tab:tcga10} to be the most plausible perturbation sites if only one perturbation exists.  
All of these top 4 sites (\textit{NFKBIA, NPEPPS, NCAM1,} and \textit{IRGM}) have been previously implicated in breast cancer.
The top-ranked gene, {\it NFKBIA}, is required for the induction and maintenance of epithelial-mesenchymal transition (EMT), and is highly relevant for metastatic processes \citep{Huber2004,Maier2010,Li2012}.

Given that genes in our list were selected for differential expression in a previous analysis of metastatic processes, we require additional validation of our results.  
We perform cross-validation to assess the predictive accuracy of the mean vector implied by each gene ranking.  
First, we divide case and control data into 10 groups of approximately equal size.  
For each fold $f$, we use 90\% of the data to estimate $\hat{\Omega}^{train}_f$ and $\hat{\mu}_{1,f}, \ldots, \hat{\mu}_{p,f}$ according to Equation~\ref{eqn:mle}.  We then predict the mean of $Y^{test}$ for each gene $j$ by taking $\hat{\Sigma}^{train}_f [0, \hat{\mu}_{j,f}, 0]^T$.  
Through this method we obtain mean-squared-error 
\begin{equation}
\mbox{MSE}_{j,f}= \frac{1}{\sum_iI(i\in f)}\sum_{i\in f}(Y_i^{test}-\hat{\Sigma}^{train}_f [0, \hat{\mu}_{j,f}, 0]^T)^2
\end{equation}
for each gene under each fold.  
We take the average of these errors to obtain a ranking of predictive ability by cross-validation, with smallest MSE indicating the best accuracy. 

\begin{table}[ht]
\centering
\caption{Top-ranked genes from multi-attribute NF analysis of TCGA methylation and gene expression data. 
$p$-values were adjusted using the Benjamini-Yekutieli method for positively correlated tests.   \label{tab:tcga10} } \vspace{.25em}
\begin{tabular}{lrrrrrrrrrr}

  \hline 
  &  \multicolumn{3}{c}{Multi-Att NF} & \multicolumn{3}{c}{Seq. NF} & \multicolumn{3}{c}{Diff. Expr.} & CV \\
  Gene & Rank & $T$ & $p_{adj}$& Rank & $T$ & $p_{adj}$ & Rank &$T^2$ & $p_{adj}$ & Rank \\ 
  \hline
  {\it NFKBIA} &   1 & 35.41 & 3.22e-06 &   1 & 35.41 & 5.53e-06 &  61 & 0.13 & 1.00e+00 &  14 \\ 
  {\it NPEPPS} &   2 & 34.84 & 3.22e-06 & 214 & 0.42 & 1.00e+00 & 215 & 1.79 & 1.00e+00 &   2 \\ 
  {\it NCAM1} &   3 & 34.30 & 3.22e-06 &  43 & 5.43 & 3.78e-01 & 241 & 0.53 & 1.00e+00 &   1 \\ 
  {\it IRGM} &   4 & 33.67 & 3.31e-06 &   2 & 30.27 & 3.61e-05 &  17 & 0.39 & 1.00e+00 &  20 \\ 
  {\it MMP13} &   5 & 28.64 & 3.27e-05 &   3 & 21.29 & 2.16e-03 &  69 & 0.50 & 1.00e+00 &  45 \\ 
  {\it MTHFD2} &   6 & 25.28 & 1.46e-04 &   9 & 11.24 & 1.07e-01 &  53 & 0.33 & 1.00e+00 &   3 \\ 
  {\it H3F3A} &   7 & 24.69 & 1.51e-04 &   4 & 17.99 & 8.39e-03 &  30 & 0.80 & 1.00e+00 &   9 \\ 
  {\it LGALS9} &   8 & 24.64 & 1.51e-04 &  15 & 9.29 & 1.61e-01 &  16 & 2.21 & 1.00e+00 &   7 \\ 
  {\it GAS8} &   9 & 21.12 & 7.79e-04 &   7 & 13.49 & 4.57e-02 &  22 & 1.14 & 1.00e+00 &  24 \\ 
  {\it UCP2} &  10 & 20.12 & 1.06e-03 &   8 & 11.83 & 9.15e-02 & 101 & 5.47 & 1.00e+00 &  27 \\ 
   \hline
\end{tabular}
\end{table}

Rankings obtained by this cross-validation procedure show a correlation with rankings from multi-attribute NF (Figure~\ref{fig:tcgastats}, bottom panel). 
This supports that the identified genes are phenotypically important elements in the network of genes under consideration, and lends credence to their identification as potential perturbation sites.

We also show ranking from  a joint differential expression analysis using Hotelling's $T^2$ test in Figure~\ref{fig:tcgastats} and Table~\ref{tab:tcga10}.  
We note that after multiple comparisons adjustments, no genes show statistically significant differential expression (only 4.8\% of our genes have a raw p-value lower than 0.05).
While a handful of top-ranked genes according to multi-attribute NF are also ranked relatively highly in differential expression results (\textit{IRGM}, \textit{LGALS9}, and \textit{GAS8}), many of our other top genes do not show strong evidence of differential expression (extended results in Supplementary Materials).
By leveraging knowledge of the interactions between genes in our data set, we are able to observe an effect from a highly relevant gene that we would not have detected by performing a multi-omic differential expression analysis.

Considering the possibility of multiple perturbations, we also performed the sequential multi-attribute NF procedure as described in Section~\ref{ss:mttest}.  
At each step, the node with the largest test statistic in the previous step is conditioned on as a nonzero portion of the mean vector, and testing is performed to ascertain whether additional nodes are nonzero.  
In the second panel of Figure~\ref{fig:tcgastats}, we see that after adjusting for a perturbation at {\it NFKBIA}, the next most likely perturbation sites is {\it IRGM}, and then {\it MMP13}, originally ranked as the fourth and fifth most likely perturbation sites respectively. 
{\it NPEPPS}, originally the second-highest ranked site, is not considered to be a likely secondary perturbation site.  
This is in agreement with known biological results; many more citations exist linking {\it NFKBIA} to metastatic processes than do for {\it NPEPPS}.  

\section{Remarks}
The multi-attribute network filtering methodology does suffer from some limitations.  
It relies upon the assumption that the network structure encoded in $\Omega$ does not vary between the control data and the case data. 
As such, this method is likely best suited to experimental settings in which it may be plausible to believe under investigator-limited 
perturbations that the underlying network relationships are fairly similar between case and control settings.

The framework here also depends upon multivariate Gaussian distributions for all data types. 
An extension of this network filtering procedure to non-Gaussian distributions would enable inclusion of additional phenotypes, such as SNP and CNV data.  
This extension has not been undertaken even in the univariate case thus far, but semi-parametric copula methods (such as those by \cite{Liu2012}) show promise for the network estimation portion of this problem.  

As is always a concern with large network models, computational costs in estimation of $\Omega$ may be prohibitive.
This is particularly the case in recovery of large, densely connected networks.  
As noted by \cite{Kolar2014}, the block gradient descent algorithm employed here performs most efficiently when the graph can be separated into smaller connected components (as a rough guide, we recommend use of this algorithm when the largest connected component has fewer than $100$ joint nodes).  
If estimation of the block-precision matrix is infeasible, use of a separated estimation procedure with network filtering, such as the joint graphical lasso \citep{Guo2011,Danaher2013}, may still be employed.  
This is expected to yield a large performance improvement over differential expression procedures, and potentially a smaller additional improvement over an analysis of a single attribute. 

Our work shows that if cross-attribute interactions are present in the data, benefits from data integration are strongest when these interactions are explicitly modeled.  
Across all tested network settings, the multi-attribute NF procedure provides better detection of perturbation sites than any single-attribute method, or multi-attribute method that ignored the network structure.  
In addition, we found that there were substantial gains to be had from a network-filtering based ranking on a single attribute  alone compared with differential expression-- it easily outperformed Hotelling's $T^2$ statistic, and provided a greater chance of an ideal identification than SSEM-lasso. 
The results in this paper underscore the need to take network effects into account when working with bioinformatic data, and offers a statistically principled method for a truly integrative analysis of multi-attribute data for better understanding cellular mechanism-of-action.

\backmatter
\section*{Supplementary Materials}
Supplementary materials are available with this paper at the Biometrics website on Wiley Online Library.  These include links to relevant software (an R package for these methods, TCGA data and processing scripts, as well as a pipeline for the simulations study), the proofs referenced in Sections ~\ref{ss:mttest} and~\ref{ss:accuracy}, and additional simulation results.

\section*{Acknowledgements}
The authors thank Mladen Kolar for sharing code for the optimization algorithm used in this paper, and to Supriya Sharma for her expertise in biological interactions.  
Thanks also to Daniel Lancour and Alexander Blocker for their help in processing data from TCGA.  
This research was supported in part by a National Institutes of Health training grant to Boston University (T32 AG000221) and AFOSR award 12RSL042. 
The computational work reported on in this paper was performed on the Shared Computing Cluster which is administered by Boston University's Research Computing Services.  (\url{www.bu.edu/tech/support/research/}).\vspace*{-8pt}

\bibliographystyle{biom} \bibliography{griffin_biometrics_bib.bib}

\begin{thebibliography}{}

\bibitem[\protect\citeauthoryear{Afzal, Mussa, Turner, Bender, and Glen}{Afzal
  et~al.}{2014}]{Afzal2014}
Afzal, A.~M., Mussa, H.~Y., Turner, R.~E., Bender, A., and Glen, R.~C. (2014).
\newblock Target fishing: A single-label or multi-label problem?
\newblock {\em arXiv preprint arXiv:1411.6285} .

\bibitem[\protect\citeauthoryear{Baldessari}{Baldessari}{1967}]{baldessari1967distribution}
Baldessari, B. (1967).
\newblock The distribution of a quadratic form of normal random variables.
\newblock {\em The Annals of Mathematical Statistics} pages 1700--1704.

\bibitem[\protect\citeauthoryear{Benjamini and Yekutieli}{Benjamini and
  Yekutieli}{2001}]{Benjamini2001}
Benjamini, Y. and Yekutieli, D. (2001).
\newblock The control of the false discovery rate in multiple testing under
  dependency.
\newblock {\em Annals of {S}tatistics} pages 1165--1188.

\bibitem[\protect\citeauthoryear{Bordbar, Mo, Nakayasu, Schrimpe-Rutledge, Kim,
  Metz, Jones, Frank, Smith, and Peterson}{Bordbar et~al.}{2012}]{Bordbar2012}
Bordbar, A., Mo, M.~L., Nakayasu, E.~S., Schrimpe-Rutledge, A.~C., Kim, Y.-M.,
  Metz, T.~O., Jones, M.~B., Frank, B.~C., Smith, R.~D., and Peterson, S.~N.
  (2012).
\newblock Model-driven multi-omic data analysis elucidates metabolic
  immunomodulators of macrophage activation.
\newblock {\em Molecular {S}ystems {B}iology} {\bf 8,} 558.

\bibitem[\protect\citeauthoryear{{Cancer Genome Atlas Network}}{{Cancer Genome
  Atlas Network}}{2012}]{TCGA2012}
{Cancer Genome Atlas Network} (2012).
\newblock Comprehensive molecular portraits of human breast tumours.
\newblock {\em Nature} {\bf 490,} 61--70.

\bibitem[\protect\citeauthoryear{Chen and Chen}{Chen and Chen}{2008}]{Chen2008}
Chen, J. and Chen, Z. (2008).
\newblock Extended {B}ayesian information criteria for model selection with
  large model spaces.
\newblock {\em Biometrika} {\bf 95,} 759--771.

\bibitem[\protect\citeauthoryear{Cosgrove, Zhou, Gardner, and
  Kolaczyk}{Cosgrove et~al.}{2008}]{Cosgrove2008}
Cosgrove, E.~J., Zhou, Y., Gardner, T.~S., and Kolaczyk, E.~D. (2008).
\newblock Predicting gene targets of perturbations via network-based filtering
  of m{RNA} expression compendia.
\newblock {\em Bioinformatics} {\bf 24,} 2482--2490.

\bibitem[\protect\citeauthoryear{Cressie}{Cressie}{1993}]{Cressie1993}
Cressie, N. (1993).
\newblock {\em Statistics for {S}patial {D}ata: {W}iley {S}eries in
  {P}robability and {S}tatistics}.
\newblock Wiley-Interscience New York.

\bibitem[\protect\citeauthoryear{Csermely, Korcsm{\'a}ros, Kiss, London, and
  Nussinov}{Csermely et~al.}{2013}]{Csermely2013}
Csermely, P., Korcsm{\'a}ros, T., Kiss, H.~J., London, G., and Nussinov, R.
  (2013).
\newblock Structure and dynamics of molecular networks: A novel paradigm of
  drug discovery: A comprehensive review.
\newblock {\em Pharmacology \& {T}herapeutics} {\bf 138,} 333--408.

\bibitem[\protect\citeauthoryear{Danaher, Wang, and Witten}{Danaher
  et~al.}{2013}]{Danaher2013}
Danaher, P., Wang, P., and Witten, D.~M. (2013).
\newblock The joint graphical lasso for inverse covariance estimation across
  multiple classes.
\newblock {\em Journal of the Royal Statistical Society: Series B (Statistical
  Methodology)} .

\bibitem[\protect\citeauthoryear{Dempster}{Dempster}{1972}]{Dempster1972}
Dempster, A.~P. (1972).
\newblock Covariance selection.
\newblock {\em Biometrics} pages 157--175.

\bibitem[\protect\citeauthoryear{di~Bernardo, Thompson, Gardner, Chobot,
  Eastwood, Wojtovich, Elliott, Schaus, and Collins}{di~Bernardo
  et~al.}{2005}]{diBernardo2005}
di~Bernardo, D., Thompson, M.~J., Gardner, T.~S., Chobot, S.~E., Eastwood,
  E.~L., Wojtovich, A.~P., Elliott, S.~J., Schaus, S.~E., and Collins, J.~J.
  (2005).
\newblock Chemogenomic profiling on a genome-wide scale using
  reverse-engineered gene networks.
\newblock {\em Nature {B}iotechnology} {\bf 23,} 377--383.

\bibitem[\protect\citeauthoryear{Du, Zhang, Huang, Jafari, Kibbe, Hou, and
  Lin}{Du et~al.}{2010}]{Du2010}
Du, P., Zhang, X., Huang, C.-C., Jafari, N., Kibbe, W.~A., Hou, L., and Lin,
  S.~M. (2010).
\newblock Comparison of {B}eta-value and {M}-value methods for quantifying
  methylation levels by microarray analysis.
\newblock {\em BMC {B}ioinformatics} {\bf 11,} 587.

\bibitem[\protect\citeauthoryear{{ENCODE Project Consortium}}{{ENCODE Project
  Consortium}}{2004}]{Encode2004}
{ENCODE Project Consortium} (2004).
\newblock The {ENCODE} ({ENC}yclopedia {O}f {DNA} {E}lements) {P}roject.
\newblock {\em Science} {\bf 306,} 636--640.

\bibitem[\protect\citeauthoryear{Fournier, Paulson, Pavelka, Mosley, Gaudenz,
  Bradford, Glynn, Li, Sardiu, Fleharty, et~al\mbox{.}}{Fournier
  et~al.}{2010}]{Fournier2010}
Fournier, M.~L., Paulson, A., Pavelka, N., Mosley, A.~L., Gaudenz, K.,
  Bradford, W.~D., Glynn, E., Li, H., Sardiu, M.~E., Fleharty, B., et~al.
  (2010).
\newblock Delayed correlation of {mRNA} and protein expression in
  rapamycin-treated cells and a role for {GGC1} in cellular sensitivity to
  rapamycin.
\newblock {\em Molecular \& {C}ellular {P}roteomics} {\bf 9,} 271--284.

\bibitem[\protect\citeauthoryear{Gotzmann, Fischer, Zojer, Mikula, Proell,
  Huber, Jechlinger, Waerner, Weith, Beug, et~al\mbox{.}}{Gotzmann
  et~al.}{2006}]{Gotzmann2006}
Gotzmann, J., Fischer, A., Zojer, M., Mikula, M., Proell, V., Huber, H.,
  Jechlinger, M., Waerner, T., Weith, A., Beug, H., et~al. (2006).
\newblock A crucial function of pdgf in tgf-$\beta$-mediated cancer progression
  of hepatocytes.
\newblock {\em Oncogene} {\bf 25,} 3170--3185.

\bibitem[\protect\citeauthoryear{Guo, Levina, Michailidis, and Zhu}{Guo
  et~al.}{2011}]{Guo2011}
Guo, J., Levina, E., Michailidis, G., and Zhu, J. (2011).
\newblock Joint estimation of multiple graphical models.
\newblock {\em Biometrika} {\bf 98,} 1--15.

\bibitem[\protect\citeauthoryear{Holland, Laskey, and Leinhardt}{Holland
  et~al.}{1983}]{Holland1983}
Holland, P.~W., Laskey, K.~B., and Leinhardt, S. (1983).
\newblock Stochastic blockmodels: First steps.
\newblock {\em Social networks} {\bf 5,} 109--137.

\bibitem[\protect\citeauthoryear{Huber, Azoitei, Baumann, Gr{\"u}nert, Sommer,
  Pehamberger, Kraut, Beug, and Wirth}{Huber et~al.}{2004}]{Huber2004}
Huber, M.~A., Azoitei, N., Baumann, B., Gr{\"u}nert, S., Sommer, A.,
  Pehamberger, H., Kraut, N., Beug, H., and Wirth, T. (2004).
\newblock Nf-$\kappa$b is essential for epithelial-mesenchymal transition and
  metastasis in a model of breast cancer progression.
\newblock {\em The Journal of clinical investigation} {\bf 114,} 569--581.

\bibitem[\protect\citeauthoryear{Kanehisa and Goto}{Kanehisa and
  Goto}{2000}]{KEGG2000}
Kanehisa, M. and Goto, S. (2000).
\newblock {KEGG}: {K}yoto encyclopedia of genes and genomes.
\newblock {\em Nucleic acids research} {\bf 28,} 27--30.

\bibitem[\protect\citeauthoryear{Kolar, Liu, and Xing}{Kolar
  et~al.}{2014}]{Kolar2014}
Kolar, M., Liu, H., and Xing, E.~P. (2014).
\newblock Graph estimation from multi-attribute data.
\newblock {\em The Journal of Machine Learning Research} {\bf 15,} 1713--1750.

\bibitem[\protect\citeauthoryear{Lecca and Priami}{Lecca and
  Priami}{2013}]{Lecca2013}
Lecca, P. and Priami, C. (2013).
\newblock Biological network inference for drug discovery.
\newblock {\em Drug {D}iscovery {T}oday} {\bf 18,} 256--264.

\bibitem[\protect\citeauthoryear{Lee, Topper, Hubler, Hose, Wenger, Coon, and
  Gasch}{Lee et~al.}{2011}]{Lee2011}
Lee, M., Topper, S.~E., Hubler, S.~L., Hose, J., Wenger, C.~D., Coon, J.~J.,
  and Gasch, A.~P. (2011).
\newblock A dynamic model of proteome changes reveals new roles for transcript
  alteration in yeast.
\newblock {\em Molecular {S}ystems {B}iology} {\bf 7,}.

\bibitem[\protect\citeauthoryear{Li and Dewey}{Li and Dewey}{2011}]{Li2011}
Li, B. and Dewey, C.~N. (2011).
\newblock {RSEM: accurate transcript quantification from RNA-seq data with or
  without a reference genome}.
\newblock {\em BMC bioinformatics} {\bf 12,} 323.

\bibitem[\protect\citeauthoryear{Li, Xia, Huo, Lim, Wu, Hsu, Chao, Yamaguchi,
  Yang, Ding, et~al\mbox{.}}{Li et~al.}{2012}]{Li2012}
Li, C.-W., Xia, W., Huo, L., Lim, S.-O., Wu, Y., Hsu, J.~L., Chao, C.-H.,
  Yamaguchi, H., Yang, N.-K., Ding, Q., et~al. (2012).
\newblock Epithelial--mesenchymal transition induced by tnf-$\alpha$ requires
  nf-$\kappa$b--mediated transcriptional upregulation of twist1.
\newblock {\em Cancer research} {\bf 72,} 1290--1300.

\bibitem[\protect\citeauthoryear{Liu, Han, Yuan, Lafferty, Wasserman,
  et~al\mbox{.}}{Liu et~al.}{2012}]{Liu2012}
Liu, H., Han, F., Yuan, M., Lafferty, J., Wasserman, L., et~al. (2012).
\newblock High-dimensional semiparametric {G}aussian copula graphical models.
\newblock {\em The Annals of Statistics} {\bf 40,} 2293--2326.

\bibitem[\protect\citeauthoryear{Ma and Zhao}{Ma and Zhao}{2012}]{Ma2012}
Ma, H. and Zhao, H. (2012).
\newblock i{F}ad: an integrative factor analysis model for drug-pathway
  association inference†.
\newblock {\em Bioinformatics} {\bf 28,} 1911--1918.

\bibitem[\protect\citeauthoryear{MacNeil, Johnson, Li, Piccolo, and
  Bild}{MacNeil et~al.}{2015}]{MacNeil2015}
MacNeil, S.~M., Johnson, W.~E., Li, D.~Y., Piccolo, S.~R., and Bild, A.~H.
  (2015).
\newblock Inferring pathway dysregulation in cancers from multiple types of
  omic data.
\newblock {\em Genome {M}edicine} {\bf 7,} 1--12.

\bibitem[\protect\citeauthoryear{Maier, Schmidt-Stra{\ss}burger, Huber,
  Wiedemann, Beug, and Wirth}{Maier et~al.}{2010}]{Maier2010}
Maier, H.~J., Schmidt-Stra{\ss}burger, U., Huber, M.~A., Wiedemann, E.~M.,
  Beug, H., and Wirth, T. (2010).
\newblock Nf-$\kappa$b promotes epithelial--mesenchymal transition, migration
  and invasion of pancreatic carcinoma cells.
\newblock {\em Cancer letters} {\bf 295,} 214--228.

\bibitem[\protect\citeauthoryear{Pham, Christadore, Schaus, and Kolaczyk}{Pham
  et~al.}{2011}]{Pham2011}
Pham, L., Christadore, L., Schaus, S., and Kolaczyk, E.~D. (2011).
\newblock Network-based prediction for sources of transcriptional dysregulation
  using latent pathway identification analysis.
\newblock {\em Proceedings of the National Academy of Sciences} {\bf 108,}
  13347--13352.

\bibitem[\protect\citeauthoryear{Szklarczyk, Franceschini, Kuhn, Simonovic,
  Roth, Minguez, Doerks, Stark, Muller, Bork, et~al\mbox{.}}{Szklarczyk
  et~al.}{2011}]{Szklarczyk2011}
Szklarczyk, D., Franceschini, A., Kuhn, M., Simonovic, M., Roth, A., Minguez,
  P., Doerks, T., Stark, M., Muller, J., Bork, P., et~al. (2011).
\newblock The {STRING} database in 2011: functional interaction networks of
  proteins, globally integrated and scored.
\newblock {\em Nucleic {A}cids {R}esearch} {\bf 39,} D561--D568.

\bibitem[\protect\citeauthoryear{Tan}{Tan}{1977}]{tan1977distribution}
Tan, W. (1977).
\newblock On the distribution of quadratic forms in normal random variables.
\newblock {\em Canadian Journal of Statistics} {\bf 5,} 241--250.

\bibitem[\protect\citeauthoryear{Tsavachidou-Fenner, Tannir, Tamboli, Liu,
  Petillo, Teh, Mills, and Jonasch}{Tsavachidou-Fenner
  et~al.}{2010}]{Tsavachidou2010}
Tsavachidou-Fenner, D., Tannir, N., Tamboli, P., Liu, W., Petillo, D., Teh, B.,
  Mills, G., and Jonasch, E. (2010).
\newblock Gene and protein expression markers of response to combined
  antiangiogenic and epidermal growth factor targeted therapy in renal cell
  carcinoma.
\newblock {\em Annals of {O}ncology} {\bf 21,} 1599--1606.

\bibitem[\protect\citeauthoryear{Varambally, Yu, Laxman, Rhodes, Mehra,
  Tomlins, Shah, Chandran, Monzon, Becich, et~al\mbox{.}}{Varambally
  et~al.}{2005}]{Varambally2005}
Varambally, S., Yu, J., Laxman, B., Rhodes, D.~R., Mehra, R., Tomlins, S.~A.,
  Shah, R.~B., Chandran, U., Monzon, F.~A., Becich, M.~J., et~al. (2005).
\newblock Integrative genomic and proteomic analysis of prostate cancer reveals
  signatures of metastatic progression.
\newblock {\em Cancer {C}ell} {\bf 8,} 393--406.

\bibitem[\protect\citeauthoryear{Wang, Singh, Zeng, Coleman, Huang, Savich, He,
  Mieczkowski, Grimm, Perou, et~al\mbox{.}}{Wang et~al.}{2010}]{Wang2010}
Wang, K., Singh, D., Zeng, Z., Coleman, S.~J., Huang, Y., Savich, G.~L., He,
  X., Mieczkowski, P., Grimm, S.~A., Perou, C.~M., et~al. (2010).
\newblock {MapSplice: accurate mapping of RNA-seq reads for splice junction
  discovery}.
\newblock {\em Nucleic acids research} {\bf 38,} e178--e178.

\bibitem[\protect\citeauthoryear{Yang and Kolaczyk}{Yang and
  Kolaczyk}{2010}]{Yang2010}
Yang, S. and Kolaczyk, E.~D. (2010).
\newblock Target detection via network filtering.
\newblock {\em {IEEE} {T}ransactions on {I}nformation {T}heory} {\bf 56,}
  2502--2515.

\bibitem[\protect\citeauthoryear{Zhang, Liu, Li, Shen, Laird, and Zhou}{Zhang
  et~al.}{2012}]{Zhang2012}
Zhang, S., Liu, C.-C., Li, W., Shen, H., Laird, P.~W., and Zhou, X.~J. (2012).
\newblock Discovery of multi-dimensional modules by integrative analysis of
  cancer genomic data.
\newblock {\em Nucleic {A}cids {R}esearch} page gks725.

\end{thebibliography}



\label{lastpage}

\end{document}


\maketitle

\section{Software}
An R package {\bf mapggm} containing methods for multi-attribute network estimation and perturbation detection is available at 
\url{https://github.com/paulajgriffin/mapggm}.  To use this package, install the {\bf devtools} package from CRAN and run:

\begin{Verbatim}
library(devtools)
install_github('paulajgriffin/mapggm')
\end{Verbatim}

\noindent
Other supplementary files, including the simulation pipeline and TCGA scripts/processed data, may be found at \url{https://github.com/paulajgriffin/mapggm\_supplemental}.

\section{Properties of sequential tests}
We prove the following theorems, presented in Section 3.2 of the paper.

\begin{theorem}{\label{th:expnest}}
Given a set of nodes already found to have nonzero mean in steps $1, \ldots s$, consider testing for a perturbation at an additional node $i$ in step $s+1$.  Denote the indices in $Z=\Omega Y$ corresponding to the nodes found in steps $1, \ldots, s$ as $S$. 

We can write the expected difference between the original test statistic and the test statistic adjusted for perturbations in $S$ as 
\begin{equation*}
E(T_i - T^{[s+1]}_i) = \mu_i^T\left(\Sigma_{i,S}\Sigma_{S,i} \right) \mu_i  + 
						 2 \mu_i^T \left(\Sigma_{i,S} \right) \mu_S + 
						 \mu_S^T \left(\Sigma_{S,i} \Sigma_{i,S} \right) \mu_S
						\enskip .
\end{equation*}
In the special case that $\mu_i = 0$, 
\begin{align*}
E(T_i - T^{[s+1]}_i | \mu_i = 0 ) \geq 0 \enskip.
\end{align*}

\end{theorem}

\noindent
{\bf Proof of Theorem~\ref{th:expnest}.}\quad First, recall the form of $T_j$, the unadjusted test statistic for testing the alternative hypothesis that $\mu_j \neq 0$ and $\mu_{(-j)}=0$ against a null of  $\mu=0$.  We can rewrite the test statistic in terms of the reordered unfiltered data $y = \left[y_j, y_{-(j)}\right]^T$ and obtain
\begin{align*}
T_j &= n \left(\bar{z}^T \Sigma \bar{z} - \bar{z}_{(-j)}^T \left(\Sigma_{(-j),(-j)}- \Sigma_{(-j),j}\Sigma_{j,j}^{-1}\Sigma_{j,(-j)}\right)\bar{z}_{(-j)} \right) \\
&=  n  \left( \bar{z}^T \Sigma \bar{z}- 
	\bar{z}^T\left( \begin{array}{cc}
		0 &  0 \\
		0 & \Sigma_{(-j),(-j)}- \Sigma_{(-j),j}\Sigma_{j,j}^{-1}\Sigma_{j,(-j)} \\
		\end{array}  \right) \bar{z}
\right)	\\
&=   n  \left( \bar{y}^T \Omega \Sigma \Omega \bar{z}- 
	\bar{y}^T \Omega \left( \begin{array}{cc}
		0 &  0 \\
		0 & \Omega_{(-j),(-j)}^{-1}\\
		\end{array}  \right) \Omega \bar{y}
\right)	\\
&= n  \left( \bar{y}^T \Omega \bar{z}- 
	\bar{y}^T \left( \begin{array}{cc}
		\Omega_{j,(-j)}\Omega_{(-j),(-j)}^{-1} \Omega_{(-j),j} &  \Omega_{j,(-j)} \\
		\Omega_{(-j),j} & \Omega_{(-j),(-j)} \\
		\end{array}  \right)  \bar{y}
\right)	\\
&= n  \left( 
	\bar{y}^T  \left( \begin{array}{cc}
		\Omega_{j,j} - \Omega_{j,(-j)}\Omega_{(-j),(-j)}^{-1} \Omega_{(-j),j} &  0 \\
		0 & 0 \\
		\end{array}  \right) \bar{y}
\right) \\
&= n  \left( 
	\bar{y}^T  \left( \begin{array}{cc}
		\Sigma_{j,j}^{-1} &  0 \\
		0 & 0 \\
		\end{array}  \right) \bar{y}
\right)
\enskip.
\end{align*}

The mean of the unfiltered data has distribution $\bar{y} \sim N (\Sigma \mu, \Sigma/n)$.  Taking the expectation of our test statistic $T_j$, we obtain

\begin{align*}
E(T_j) =& n E(\bar{y})^T \left( \begin{array}{cc}
		\Sigma_{j,j}^{-1} &  0 \\
		0 & 0 \\
		\end{array}  \right) E(\bar{y}) + 
		\mbox{Tr} \left( 
		\left( \begin{array}{cc}
		\Sigma_{j,j}^{-1} &  0 \\
		0 & 0 \\
		\end{array}  \right) \Sigma
		\right)\\
=& n 
\left( \begin{array}{c}
		\Sigma_{j,j}\mu_j + \Sigma_{j,(-j)}\mu_{(-j)}  \\
		\Sigma_{(-j),j}\mu_j + \Sigma_{(-j),(-j)}\mu_{(-j)} \\
		\end{array}  \right)^T
\left( \begin{array}{cc}
		\Sigma_{j,j}^{-1} &  0 \\
		0 & 0 \\
		\end{array}  \right) 
\left( \begin{array}{c}
		\Sigma_{j,j}\mu_j + \Sigma_{j,(-j)}\mu_{(-j)}  \\
		\Sigma_{(-j),j}\mu_j + \Sigma_{(-j),(-j)}\mu_{(-j)} \\
		\end{array}  \right)		
		 \\ 
		 &+ 
		\mbox{Tr} \left( 
		\left( \begin{array}{cc}
		I &  \Sigma_{j,j}^{-1}\Sigma_{j,(-j)} \\
		0 & 0 \\
		\end{array}  \right) 
		\right)\\
=& n \left(\mu_j^T \Sigma_{j,j} \mu_j  + 
	2\mu_j^T\Sigma_{j,(-j)}\mu_{(-j)} +
	\mu_{(-j)}^T \Sigma_{(-j),j}  \Sigma_{j,(-j)} \mu_{(-j)}
	\right) + k_j	\enskip,
\end{align*}
where $k_j$ indicates the number of attributes for node $j$.

Denote the indices in $Z=\Omega Y$ corresponding to the nodes found in steps $1, \ldots, s$ as $S$, and the indices corresponding to the node currently under consideration as $i$.  
Denote all other indices $X$. 

In the sequential testing procedure, we test the alternative hypothesis of $\mu_i \neq 0$, $\mu_S \neq 0$, $\mu_X =0$ against the null that $\mu_S \neq 0$, $\mu_i=0$, $\mu_X=0$.  We can write the adjusted test statistic $T_i^{[s+1]}$ as the difference of two unadjusted test statistics

\begin{equation}
T_i^{[s+1]} = T_{(i,S)} - T_S \enskip.
\end{equation}

We are interested in $E(T_i - T_i^{[s+1]})$, the expected difference between the original and adjusted test statistic. 
\begin{alignat*}{2}
E(T_i - T_i^{[s+1]})  =& &&E(T_i) + E(T_S)  - E(T_{(i,S)}) \\
  =& && n \left( 
 \mu_i^T \Sigma_{ii} \mu_i + 2\mu_i^T \Sigma_{i,S} \mu_s + 2 \mu_i^T \Sigma_{i,X} \mu_X + \mu_S^T \Sigma_{S,i} \Sigma_{i,S} \mu_S + \mu_X^T \Sigma_{X,i}\Sigma_{i,X} \mu_X
 \right) + k_i \\
 &+ &&n \left( \mu_S^T \Sigma_{S,S} \mu_S + 2 \mu_S^T \Sigma_{S,i} \mu_i + 
 				2\mu_S^T \Sigma_{S,X} \mu_X+
 				\mu_i^T \Sigma_{i,S}\Sigma_{S,i} \mu_i + \right. \\
& && \left. \mu_X^T \Sigma_{X,S} \Sigma_{S,X} \mu_X
		\right) + \sum_{j \in S}k_j 	\\
&- &&n \left( 
\mu_i^T \Sigma_{i,i} \mu_i + 2\mu_i^T \Sigma_{i,S} \mu_S + \mu_S^T \Sigma_{S,S} \mu_S
 + 2\mu_i^T \Sigma_{i,X}\mu_X + 2\mu_S^T \Sigma_{S,X} \mu_X\right. \\
& &&+ \left. 
 \mu_X^T \Sigma_{X,i}\Sigma_{i,X} \mu_X + \mu_X^T \Sigma_{X,S} \Sigma_{S,X} \mu_x
 \right)	+ \left(k_i + 	\sum_{j \in S}k_j \right)			
\end{alignat*}
\noindent
By gathering common terms, we obtain 
\begin{equation*}
E(T_i - T^{[s+1]}_i) = \mu_i^T\left(\Sigma_{i,S}\Sigma_{S,i} \right) \mu_i  + 
						 2 \mu_i^T \left(\Sigma_{i,S} \right) \mu_S + 
						 \mu_S^T \left(\Sigma_{S,i} \Sigma_{i,S} \right) \mu_S
						\enskip .
\end{equation*}
\noindent
In the special case that $\mu_i = 0$ (no perturbation exists at the node under consideration),
\begin{align*}
E(T_i - T^{[s+1]}_i | \mu_i = 0 ) =  \mu_S^T \left(\Sigma_{S,i} \Sigma_{i,S} \right) \mu_S \geq 0 \enskip,
\end{align*}
since $\left(\Sigma_{S,i} \Sigma_{i,S}\right)$ is by definition positive semi-definite.


\clearpage
\begin{theorem}{\label{th:nested}}
Under the same conditions outlined in the general case of Theorem~\ref{th:expnest}, if $\Sigma_{i,S} = 0$, then
\begin{equation*}
T^{[s+1]}_i= T_i \enskip .
\end{equation*}

\end{theorem}

\noindent
{\bf Proof of Theorem~\ref{th:nested}.}\quad 
Denote the indices in $Z=\Omega Y$ corresponding to the nodes found in steps $1, \ldots, s$ as $S$, and the indices corresponding to the node currently under consideration as $i$.  Denote all other indices $X$.

For any $\Sigma_{S,i}$ we can write the test statistic $T_i$ for the unconditional test as
\begin{align*}
T_i &= n(\bar{z} - \hat{\mu}_A)^T \Sigma (\bar{z} - \hat{\mu}_A) - n(\bar{z} - \hat{\mu}_0)^T \Sigma (\bar{z} - \hat{\mu}_0)\\ 
&= n\left((\bar{z} - \hat{\mu}_A) - (\bar{z} - \hat{\mu}_0) \right)^T \Sigma \left((\bar{z} - \hat{\mu}_A) + (\bar{z} - \hat{\mu}_0) \right) 
\enskip,
\end{align*}
 where $\hat{\mu}_0$ and $\hat{\mu}_A$ denote the maximum likelihood estimators for $\mu$ under the null and alternative hypothesis, respectively.  Without loss of generality, we reorder the filtered data so that $Z = \left(Z_i', Z_S', Z_X'\right)$

Following formula (7) in the main paper, for the unconditional test, we have
 \begin{align*}
\hat{\mu}_0 &= \left(\begin{array}{c} 0 \\ 0\\ 0 \\ \end{array} \right)  &\enskip,
\end{align*}
and 
\begin{align*}
\hat{\mu}_A &= \left( \begin{array}{c} 
			 \bar{z}_i+\Sigma_{i,i}^{-1}\Sigma_{i,(SX)}\bar{z}_{(SX)} \\ 
				0\\
				0 \\ \end{array}  \right) &\enskip.
\end{align*}

\noindent
Similarly, a nested likelihood ratio test that conditions on the presence of nonzero mean values for indices $S$ has the form
\begin{align*}
T_i^{[s+1]} &= n(\bar{z} - \hat{\mu}_A^{[s+1]})^T \Sigma (\bar{z} - \hat{\mu}_A^{[s+1]}) - n(\bar{z} - \hat{\mu}_0^{[s+1]})^T \Sigma (\bar{z} - \hat{\mu}_0^{[s+1]})\\ 
&= n\left((\bar{z} - \hat{\mu}_A^{[s+1]}) - (\bar{z} - \hat{\mu}_0^{[s+1]}) \right)^T \Sigma \left((\bar{z} - \hat{\mu}_A^{[s+1]}) + (\bar{z} - \hat{\mu}_0^{[s+1]}) \right) 
\enskip,
\end{align*}

\noindent
with restricted MLEs
 \begin{align*}
\hat{\mu}_0^{[s+1]} &= \left(\begin{array}{c} 
			0 \\ 
			\bar{z}_S + \Sigma_{S,S}^{-1} \Sigma_{S, (iX)} \bar{z}_{(iX)}\\ 
			0 \\ \end{array} \right)  &\enskip, \mbox{ and} \\
\hat{\mu}_A^{[s+1]} &= \left( \begin{array}{c} 
			\bar{z}_{(iS)}+\Sigma_{(iS),(iS)}^{-1}\Sigma_{(iS),X}\bar{z}_{X} \\ 
			0 \\ \end{array}  \right) &\enskip.
\end{align*}

Recall $\Sigma_{i,S}=0$ by assumption. We may rewrite $\hat{\mu}_A$, $\hat{\mu}_0^{[s+1]}$, and $\hat{\mu}_A^{[s+1]}$ as
 \begin{align*}
\hat{\mu}_A 
	&= \left( \begin{array}{c} 
			\bar{z}_i+\Sigma_{i,i}^{-1}(\Sigma_{i,S}\bar{z}_S+ \Sigma_{i,X}\bar{z}_X) \\ 
			0\\
			0 \\ \end{array}  \right) \\ 
	&= \left( \begin{array}{c} 
			\bar{z}_i+\Sigma_{i,i}^{-1}\Sigma_{i,X}\bar{z}_X \\ 
			0\\
			0 \\ \end{array}  \right) \\ 
\hat{\mu}_0^{[s+1]} 
	&= \left(\begin{array}{c} 
			0 \\ 
			\bar{z}_S + \Sigma_{S,S}^{-1} (\Sigma_{S,i} \bar{z}_i + \Sigma_{S,X}\bar{z}_X)\\ 
			0 \\ \end{array} \right)  \\ 
	&= \left(\begin{array}{c} 
			0 \\ 
			\bar{z}_S + \Sigma_{S,S}^{-1}\Sigma_{SX}\bar{z}_X\\ 
			0 \\ \end{array} \right)  \\
\hat{\mu}_A^{[s+1]} 
	&= \left( \begin{array}{c} 
			\bar{z}_{(iS)} + \Sigma_{(iS),(iS)}^{-1}\Sigma_{(iS),X}\bar{z}_{X} \\ 
		  0 \\ \end{array}  \right) \\
	&= \left( \begin{array}{c} 
		  \bar{z}_{i}+\Sigma_{i,i}^{-1}\Sigma_{i,X}\bar{z}_{X} \\  
		  \bar{z}_{S}+\Sigma_{S,S}^{-1}\Sigma_{S,X}\bar{z}_{X} \\
		  0 \\ \end{array}  \right) \enskip.
\end{align*}

\noindent
Our unadjusted test yields
\begin{align*}
T_i
	&= n\left(\left(\begin{array}{c} 
			-\Sigma_{i,i}^{-1}\Sigma_{i,X}\bar{z}_X \\ 
			\bar{z}_S\\
			\bar{z}_X \\ \end{array}\right) - 
			\left(\begin{array}{c} 
			\bar{z}_i\\ 
			\bar{z}_S\\
			\bar{z}_X\\ \end{array}\right) \right)^T 
		\Sigma 
		 \left(\left(\begin{array}{c} 
			-\Sigma_{i,i}^{-1}\Sigma_{i,X}\bar{z}_X \\ 
			\bar{z}_S\\
			\bar{z}_X \\ \end{array}\right) + 
			\left(\begin{array}{c} 
			\bar{z}_i\\ 
			\bar{z}_S\\
			\bar{z}_X\\ \end{array}\right) \right) \\
	&= n\left(\begin{array}{c} 
			-\bar{z}_i -\Sigma_{i,i}^{-1}\Sigma_{i,X}\bar{z}_X \\ 
			0 \\
			0 \\ \end{array}\right)^T 
		\Sigma 
		 \left(\begin{array}{c} 
			\bar{z}_i-\Sigma_{i,i}^{-1}\Sigma_{i,X}\bar{z}_X \\ 
			2\bar{z}_S\\
			2\bar{z}_X \\ \end{array} \right) 
\end{align*}

\noindent
By a similar process, the adjusted test statistic is
\begin{align*}
T_i^{[s+1]}
	&= n\left(\begin{array}{c} 
			-\bar{z}_i -\Sigma_{i,i}^{-1}\Sigma_{i,X}\bar{z}_X \\ 
			0 \\
			0 \\ \end{array}\right)^T 
		\Sigma 
		 \left(\begin{array}{c} 
			\bar{z}_i-\Sigma_{i,i}^{-1}\Sigma_{i,X}\bar{z}_X \\ 
			2\bar{z}_S -2\Sigma_{S,S}^{-1}\Sigma_{SX}\bar{z}_X \\
			2\bar{z}_X \\ \end{array} \right) \enskip.
\end{align*}

\noindent
Note that both of these statistics have the form
\begin{align*}
T &= n d' \Sigma a \\
  &= n \left( \begin{array}{c} 
			d_i \\ 
			d_S \\
			d_X \\ \end{array}
  \right)^T \left(
		\begin{array}{ccc}
		\Sigma_{ii}& 0& \Sigma_{iX} \\
		0& \Sigma_{SS}& \Sigma_{SX} \\
		\Sigma_{Xi}& \Sigma_{XS}& \Sigma_{XX} \\
		\end{array}
  \right) 
  \left(\begin{array}{c} 
			a_i \\ 
			a_S \\
			a_X \\ \end{array} \right)\\
&= d_i^T \left(\Sigma_{ii} a_i + \Sigma_{iX}a_X\right) 
	+ d_S^T \left(\Sigma_{SS} a_S + \Sigma_{SX}a_X\right) 
	+ d_X^T \left(\Sigma_{Xi} a_i + \Sigma_{XS} a_S + \Sigma_{XX}a_X\right) \\
&= d_i^T \left(\Sigma_{ii} a_i + \Sigma_{iX}a_X\right) 	\enskip.
\end{align*}

\noindent
In both $T_i$ and $T_i^{[s+1]}$, we have $d_i=-\bar{z}_i -\Sigma_{i,i}^{-1}\Sigma_{i,X}\bar{z}_X$, $a_i = \bar{z}_i-\Sigma_{i,i}^{-1}\Sigma_{i,X}\bar{z}_X$, and $a_X=2\bar{z}_x$. Therefore, $T_i^{[s+1]} = T_i$.

\section{Bounds on error in test statistic}
We prove the following theorem, presented in Section 3.3 of the paper.
\begin{theorem}{\label{th:errbound}}
Under the conditions above, the discrepancy $T_1-\tilde T_1$ is equal in distribution to a linear combination
of mutually independent, noncentral chisquare random variables,  
\begin{equation}
\sum_{k=1}^s a_k \chi^2_{r_k}\left(\delta_k\right) \enskip ,
\label{eq:main.decomp}
\end{equation}
where 
$$\delta_k = (n/2) \mu^T \Sigma_{\cdot 1} E_k \Sigma_{11}^{-1} \Sigma_{1\cdot}\mu \enskip .$$
Accordingly,
\begin{equation}
E\left[ T_1 - \tilde T_1\right] = tr\left(D\Sigma_{11}\right) + \frac{1}{2}n\mu^T \Sigma_{\cdot 1} D\Sigma_{1\cdot} \mu
\label{eq:mean}
\end{equation}
and
\begin{equation}
\hbox{Var}\left(T_1 - \tilde T_1\right) = 2 tr\left( (D\Sigma_{11})^2\right) 
             + 2n \mu^T \Sigma_{\cdot 1} D \Sigma_{11} D \Sigma_{1\cdot} \mu \enskip .
\label{eq:var}
\end{equation}
\end{theorem}
\medskip

\noindent
{\bf Proof of Theorem~\ref{th:errbound}.}\quad 
Begin by noting that $T_1 - \tilde T_1 = X^T D X$, where 
$$D = \Omega_{11} - \Omega_{1\cdot}\Omega_{\cdot\cdot}^{-1}\Omega_{\cdot 1} -
         \left(\tilde\Omega_{11} - \tilde\Omega_{1\cdot}\tilde\Omega_{\cdot\cdot}^{-1}\tilde\Omega_{\cdot 1} \right) \enskip ,$$
as defined in the paper, and $X$ is a multivariate normal random variable with mean 
$n^{1/2} \Sigma_{1\cdot}\mu$ and covariance $\Sigma_{11}$.  Since $D$ is symmetric and $\Sigma_{11}$ is symmetric and positive definite
(the latter by assumption), it follows from Lemma 1 of \cite{baldessari1967distribution} that $D\Sigma_{11}$ has spectral decomposition
$$D\Sigma_{11} = \sum_{k=1}^s a_k E_k\enskip ,$$
such that $\hbox{rank}(E_k) = r_k$ (corresponding to the multiplicity of the eigenvalue $a_k$) and $\sum_{k=1}^s r_k = K$.
By direct application of Theorem 1 of \cite{baldessari1967distribution}, the expression in (\ref{eq:main.decomp}) then follows.

As for the mean and variance expressions in (\ref{eq:mean}) and (\ref{eq:var}), we see that
\begin{eqnarray*}
E\left[ T_1 - \tilde T_1\right] & = & E\left[ \sum_{k=1}^s a_k \chi^2_{r_k}(\delta_k)\right] \\
     & = & \sum_{k=1}^s a_k(r_k + \delta_k) \\
     & = & \sum_{k=1}^s a_k r_k  + \sum_{k=1}^s a_k\delta_k \\
     & = & tr(D\Sigma_{11}) + \frac{n}{2} \mu^T \Sigma_{\cdot 1} D\Sigma_{1\cdot} \mu \enskip ,
\end{eqnarray*}
and similarly,
\begin{eqnarray*}
\hbox{Var}\left(T_1 - \tilde T_1\right) & = & \hbox{Var}\left(\sum_{k=1}^s a_k \chi^2_{r_k}(\delta_k)\right) \\
 & = & \sum_{k=1}^s a_k^2( 2r_k + 4\delta_k) \\
  & = & 2\sum_{k=1}^s a_k^2 r_k + 4 \sum_{k=1}^s a_k^2 \delta_k \\
  & = &   2 tr\left( (D\Sigma_{11})^2\right) 
             + 2n \mu^T \Sigma_{\cdot 1} D \Sigma_{11} D \Sigma_{1\cdot} \mu \enskip ,
\end{eqnarray*}
where we have exploited independence among the chisquare random variables in both cases.
\bigskip

The following corollary was also provided in Section 3.3 of the paper.
\begin{corollary}{\label{cor:meanvar}}
Let $||\cdot ||_2$ denote the spectral norm.  Then
$$E\left[T_1 - \tilde T_1\right] = O\left(||\Delta||_2\right) \quad\hbox{and}\quad
\hbox{Var}\left(T_1 - \tilde T_1\right) = O\left(||\Delta||^2_2\right) \enskip .$$
\end{corollary}
\medskip

\noindent {\bf Proof of Corollary~\ref{cor:meanvar}.}\quad
The statements in this corollary follow through application of bounds on the trace of matrix products and repeated application
of Cauchy-Schwartz, coupled with an appeal to the Lipschitz smoothness of the mapping between $\Omega$  and
the expression $ \Omega_{11} - \Omega_{1\cdot}\Omega_{\cdot\cdot}^{-1}\Omega_{\cdot 1}$.  The latter follows from a
straightforward Taylor series argument and the continuity of matrix inversion.

\cite{wang1986trace} establish that for two matrices $M$ and $N$, with $N$ symmetric and positive semidefinite,
that $|tr(MN)|\le ||M||_2 \, tr(N)$.  For the mean, therefore, we have that $tr(D\Sigma_{11}) \le ||D||_2\, tr(\Sigma_{11})$.  At the same time, 
$$\big\vert \frac{n}{2}\mu^T \Sigma_{\cdot 1} D\Sigma_{1\cdot} \mu \big\vert 
   \le \frac{n}{2} ||\Sigma_{1\cdot} \mu||^2_2 ||D||_2 \enskip .$$
As a result, we find that $E[T_1 - \tilde T_1] = O(||D||_2)$.

Similarly, for the variance
$$2 tr\left( (D\Sigma_{11})^2 \right) \le 2 tr\left( D^2 \Sigma^2_{11}\right) 
\le 2 ||D^2||_2 \, tr(\Sigma^2_{11}) \le 2 ||D||^2_2\, tr(\Sigma^2_{11}) \enskip ,$$
where the first inequality follows from Theorem 1 of \cite{chang1999matrix}.  Additionally,
$$|2n \mu^T \Sigma_{\cdot 1} D \Sigma_{11} D \Sigma_{1\cdot} \mu | \le
   2n ||\Sigma_{11}\mu||^2_2 ||\Sigma_{11}||_2 ||D||^2_2 \enskip .$$
Hence, $\hbox{Var}(T_1 - \tilde T_1) = O(||D||^2_2)$.

Recall that the quantity $||D||_2$ depends upon our choice of $j=1$.  In order to have a general result,
applicable to $T_j - \tilde T_j$ for all $j$, we prefer a bound in terms of the overall error $\Delta = \tilde \Omega - \Omega$.
Without loss of generality, define a function $f(\Omega) = \Omega_{11} - \Omega_{1\cdot}\Omega_{\cdot\cdot}^{-1}\Omega_{\cdot 1}$.
That this function is Lipschitz smooth is straightforward to show, as mentioned previously.  As a result,
$$||D||_2 = ||f(\Omega) - f(\tilde\Omega)||_2 \le K ||\Omega - \tilde \Omega||_2 = K ||\Delta ||_2 \enskip .$$
The results of the corollary then follow.

\section{Additional simulations}
Additional simulations are provided to demonstrate predictive ability at lower signal-to-noise (SNR) thresholds. Comparisons across values of $\rho_{in}$ and $\rho_{out}$ in the main paper were shown with $\mbox{SNR}=0.20$; these additional simulations show $\mbox{SNR}=0.10$ and $\mbox{SNR}=0.05$.

\begin{figure}[ht]
\centering
\includegraphics[width=\textwidth]{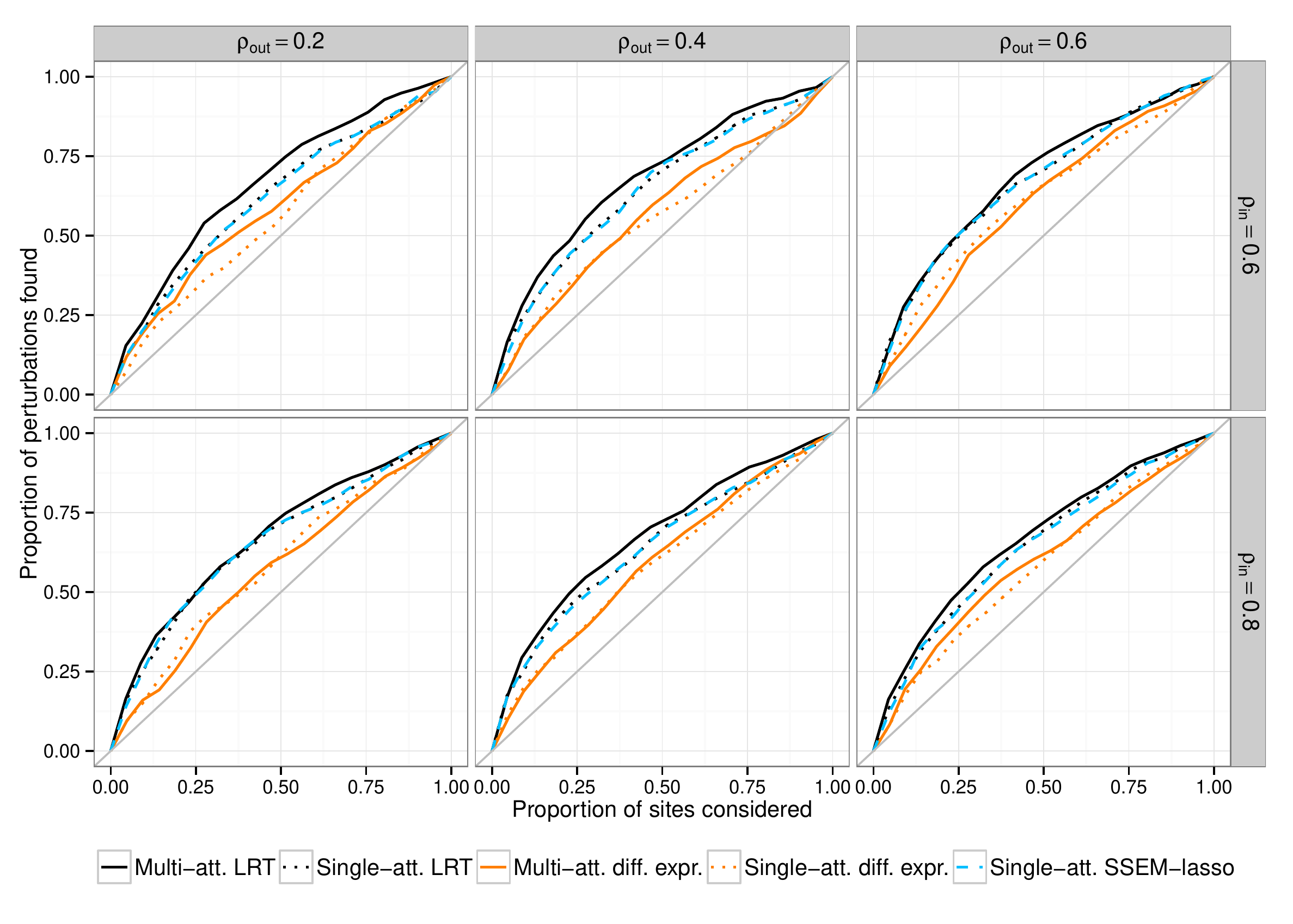}
\caption{Single-site recovery from a stochastic block model simulation with $p=20$ nodes, $n=50$ cases and controls, and $\mbox{SNR}=0.05$.\label{fig:snr10}}
\end{figure}

\begin{table}[ht]
\centering
\caption{Probability that the top-ranked site is the true perturbation site and (AUC) for simulations shown in Figure~\ref{fig:snr10}. ($\mbox{SNR}=0.10$)}
\label{tab:snr10}
\begin{tabular}{ccccccc}
  \hline
 &  & \multicolumn{2}{c}{LRT methods}  & \multicolumn{2}{c}{Differential expression} & SSEM-lasso \\ 
$\rho_{in}$ & $\rho_{out}$ & Multi-att. & Single-att. & Multi-att. & Single-att.  & Single-att.\\ 
  \hline
   0.8 & 0.2  & 0.14 (0.61) & 0.12 (0.61) & 0.10 (0.58) & 0.11 (0.55) & 0.11 (0.60) \\ 
       & 0.4  & 0.21  (0.65)& 0.14 (0.63) & 0.12 (0.59) & 0.12 (0.57) & 0.12 (0.62) \\ 
       & 0.6  & 0.16 (0.68) & 0.14 (0.63) & 0.09 (0.58) & 0.08 (0.55) & 0.14 (0.62) \\ \hline
   0.6 & 0.2  & 0.21 (0.70) & 0.15 (0.65) & 0.14 (0.64) & 0.14 (0.58) & 0.17 (0.65) \\ 
       & 0.4  & 0.09  (0.64)& 0.11 (0.64) & 0.09 (0.58) & 0.10 (0.57) & 0.11 (0.63) \\ 
       & 0.6  & 0.15 (0.67) & 0.11 (0.64) & 0.07 (0.55) & 0.09 (0.55) & 0.12 (0.63) \\ 
   \hline
\end{tabular}
\vspace{1em}
\end{table}

\begin{figure}[ht]
\centering
\includegraphics[width=1\textwidth]{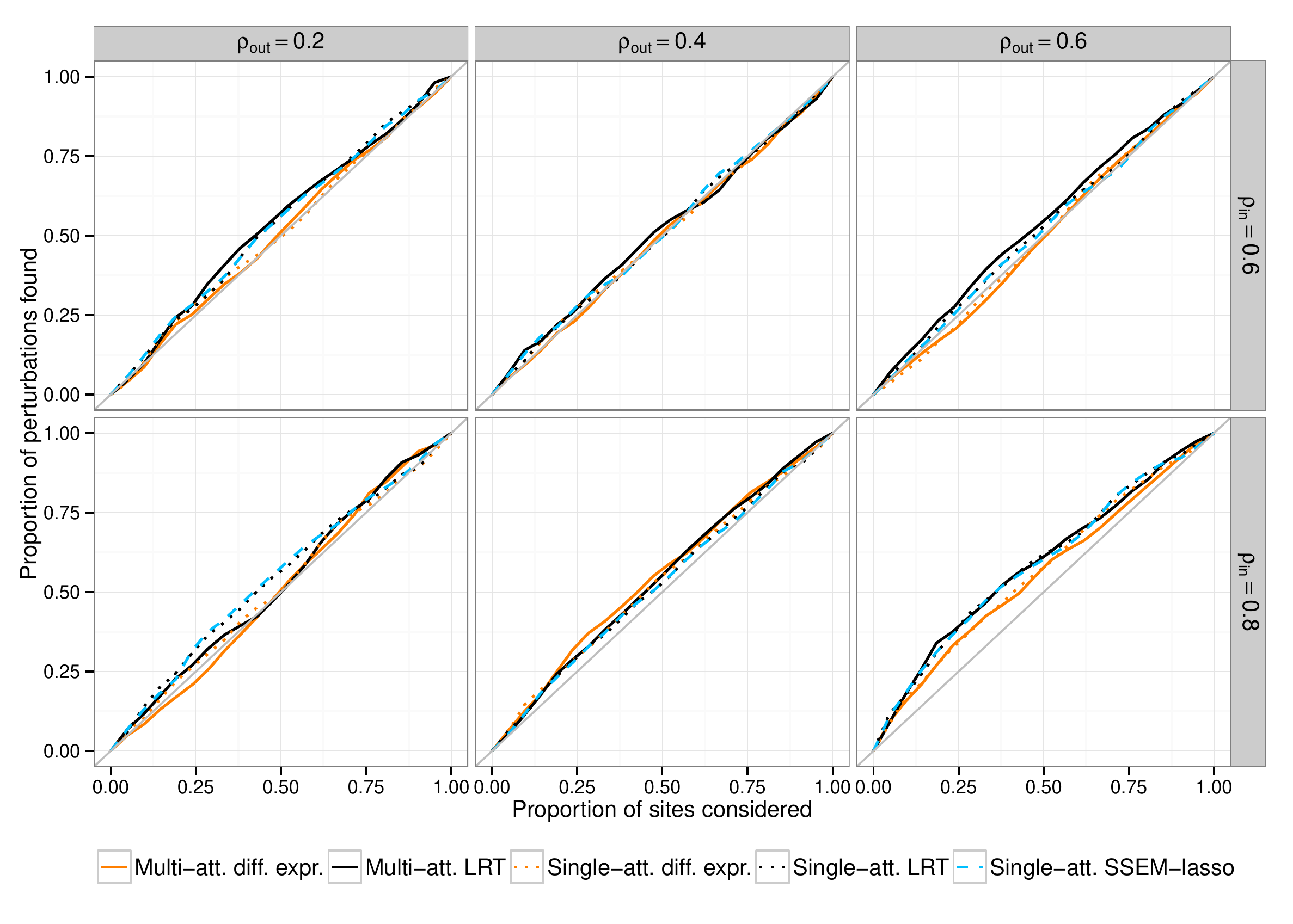}
\caption{Single-site recovery from a stochastic block model simulation with $p=20$ nodes, $n=50$ cases and controls, and $\mbox{SNR}=0.05$.}\label{fig:snr05}.
\end{figure}

\begin{table}[ht]
\centering
\caption{Probability that the top-ranked site is the true perturbation site and (AUC) for simulations shown in Figure~\ref{fig:snr05}. ($\mbox{SNR}=0.10$)}
\label{tab:snr05}
\begin{tabular}{ccccccc}
  \hline
 &  & \multicolumn{2}{c}{LRT methods}  & \multicolumn{2}{c}{Differential expression} & SSEM-lasso \\ 
$\rho_{in}$ & $\rho_{out}$ & Multi-att. & Single-att. & Multi-att. & Single-att.  & Single-att.\\ 
  \hline
0.8 & 0.2 & 0.09 (0.52) & 0.09 (0.51) & 0.05 (0.52) & 0.06 (0.52) & 0.07 (0.51) \\ 
    & 0.4 & 0.09 (0.55) & 0.08 (0.55) & 0.07 (0.54) & 0.05 (0.55) & 0.07 (0.54) \\ 
    & 0.6 & 0.07 (0.55) & 0.08 (0.54) & 0.08 (0.48) & 0.05 (0.48) & 0.07 (0.54) \\ \hline
0.6 & 0.2 & 0.10 (0.56) & 0.08 (0.57) & 0.07 (0.52) & 0.06 (0.52) & 0.09 (0.56) \\ 
    & 0.4 & 0.06 (0.55) & 0.09 (0.52) & 0.08 (0.54) & 0.07 (0.56) & 0.07 (0.52) \\ 
    & 0.6 & 0.09 (0.56) & 0.10 (0.53) & 0.07 (0.52) & 0.06 (0.52) & 0.10 (0.54) \\ 
   \hline
\end{tabular}
\vspace{1em}
\end{table}

\clearpage
\bibliographystyle{biom} \bibliography{griffin_biometrics_bib.bib}